\documentclass[a4paper,11pt]{article}
\pdfoutput=1 % if your are submitting a pdflatex (i.e. if you have
\usepackage[utf8]{inputenc}
\usepackage{jinstpub} % for details on the use of the package, please
\usepackage{booktabs}
\usepackage{subfigure}
\usepackage{changepage}
\usepackage{textcomp}
\usepackage{xcolor}
\usepackage{float}
\usepackage{upgreek}
\usepackage{siunitx}
\DeclareSIUnit{\neq}{n_{\text{eq}}\per\cm\squared}
\DeclareSIUnit{\MRad}{\mega{Rad}}
\newcommand{\micron}{\textmu\rm{m}}

\usepackage{graphicx}
\usepackage{capt-of}% or \usepackage{caption}
\usepackage{varwidth}

\title{Test beam measurement of ams H35 HV-CMOS capacitively coupled pixel sensor prototypes with high-resistivity substrate}

\author[a,1]{M.~Benoit,\note{Corresponding author.}}
\author[b]{S.~Braccini,}
\author[h]{R.~Casanova,}
\author[h]{E.~Cavallaro,}
\author[d]{H.~Chen,}
\author[d]{K.~Chen,}
\author[a]{F.A.~Di Bello,}
\author[a]{D.~Ferrere,}
\author[l]{D.~Frizzell,}
\author[a]{T.~Golling,}
\author[a]{S.~Gonzalez-Sevilla,}
\author[h]{S. Grinstein,}
\author[a]{G.~Iacobucci,}
\author[a]{M.~Kiehn,}
\author[d]{F.~Lanni,}
\author[d,e]{H.~Liu,}
\author[j]{J.~Metcalfe,}
\author[a,c]{L.~Meng,}
\author[b]{C.~Merlassino,}
\author[b]{A.~Miucci,}
\author[f]{D.~Muenstermann,}
\author[a,g]{M.~Nessi,}
\author[h]{H.~Okawa,}
\author[i]{I.~Peri\'c,}
\author[b]{M.~Rimoldi,}
\author[a,g]{B.~Risti\'c,}
\author[a]{D M S Sultan,}
\author[h]{S. Terzo,}
\author[a]{M.~Vicente Barrero Pinto,}
\author[c]{E. Vilella Figueras,}
\author[b]{M.~Weber,}
\author[b]{T.~Weston,}
\author[d]{W.~Wu,}
\author[j]{J. Xie,}
\author[d]{L.~Xu,}
\author[a]{E.~Zaffaroni,}
\author[m]{and M.~Zhang}
\affiliation[a]{D\'epartement de Physique Nucl\'eaire et Corpusculaire
  (DPNC), Université de Genève, 24 quai Ernest Ansermet 1211 Gen\`eve 4, Switzerland}
\affiliation[b]{Albert Einstein Center for Fundamental Physics and Laboratory for High Energy Physics, University of Bern, Siedlerstrasse 5, CH-3012 Bern, Switzerland}
\affiliation[c]{Department of Physics, University of Liverpool, The Oliver Lodge Laboratory, Liverpool L69 7ZE, UK}
\affiliation[d]{Brookhaven National Laboratory (BNL), P.O. Box 5000, Upton, NY 11973-5000, USA}
\affiliation[e]{Dept. of Modern Physics, University of Science and Technology of China, Hefei, Anhui 230026, China}
\affiliation[f]{Lancaster University, Physics Department, Lancaster, LA1 4YB, UK}
\affiliation[g]{European Organization for Nuclear Research (CERN), 385 route de Meyrin, 1217 Meyrin, Switzerland}
\affiliation[h]{Faculty of Pure and Applied Sciences, and CiRfSE, University of Tsukuba, Tsukuba 305-8571, Japan}
\affiliation[i]{Karlsruhe Institute of Technology (KIT), IPE, 76021 Karlsruhe, Germany}
\affiliation[j]{Argonne National Laboratory (ANL), Argonne, IL 60439, USA}
\affiliation[k]{Institut de Física d’Altes Energies (IFAE) and ICREA, The Barcelona Institute of Science and Technology, Edifici CN, UAB campus, 08193 Bellaterra (Barcelona), Spain}
\affiliation[l]{University of Oklahoma, 660 Parrington Oval, Norman, OK 73019, USA}
\affiliation[m]{University of Illinois Urbana Champaign, 1110 W Green St Loomis Laboratory, Urbana, IL 61801, USA}

\emailAdd{Mathieu.Benoit@cern.ch}

\abstract{In the context of the studies of the ATLAS High Luminosity LHC programme, radiation tolerant pixel detectors in CMOS technologies are investigated. To evaluate the effects of substrate resistivity on CMOS sensor performance, the H35DEMO demonstrator, containing different diode and amplifier designs, was  produced in ams H35 HV-CMOS technology using four different substrate resistivities spanning from \SIrange{80}{1000}{\ohm\cm}. A glueing process using a high-precision flip-chip machine was developed in order to capacitively couple the sensors to FE-I4 Readout ASIC using a thin layer of epoxy glue with good uniformity over a large surface. The resulting assemblies were measured in beam test at the Fermilab Test Beam Facilities with \SI{120}{\GeV} protons and CERN SPS H8 beamline using \SI{180}{\GeV} pions. The in-time efficiency and tracking properties measured for the different sensor types are shown to be compatible with the ATLAS ITk requirements for its pixel sensors.}


\begin{document}

\maketitle

\flushbottom

\section{Introduction}\label{sec:introduction}

The new ATLAS inner detector for the High Luminosity LHC programme,
called ITk, will require a large production of radiation
tolerant pixel detectors in order to cover the large ITk surface area. CMOS technologies, where high-voltage and high-resistivity substrate can be used, represent a promising avenue to produce large number of low cost pixel detectors, taking advantage of the large scale industrial production facilities offered by the CMOS foundries. The high resistivity of the substrate, combined with the possibility to apply bias voltage larger than \SI{120}{\V} allows the creation of a large depletion zone within the sensor substrate with a drift field sufficient to generate fast and large enough signals both before and after the irradiation of the sensors. 

Previous  small prototypes, produced in multi-project wafer projects in ams h18 technology \cite{ams} using standard low-resistivity substrate have shown good tracking performances before and after irradiation using capacitive-coupling of the sensor to its readout electronics \cite{IvanPeric,CCPD,CCPDv4-paper,CCPDV4irrad,marcos}. However, it was concluded that higher signal from high-resistivity substrate would contribute to a better timing and better radiation hardness.

To demonstrate the possibility of producing CMOS sensors using high-resistivity substrates, a prototype of large area CMOS sensor, the H35DEMO, was designed, for capacitive-coupling to a readout ASIC or standalone readout. This prototype was produced in an engineering run in ams h35 technology which first allowed the use of high-resistivity substrates. This prototype was assembled to readout ASICs using capacitive-coupling and studied in test beam campaigns to evaluate its tracking properties, measure the improvement in signal expected using higher resistivity substrate and demonstrate the feasibility of producing large area sensors of the order of \si{\cm\squared} as required for ATLAS ITk project. However, poor performances are expected from the ams h35 $350 nm$ technology after irradiation and the capacitively coupled prototypes were not design to withstand ATLAS ITk specifications in terms of radiation. They have therefore been characterised un-irradiated.

\section{The H35DEMO demonstrator chip}

\begin{figure}[h]
\center
  \includegraphics[width=0.7\textwidth]{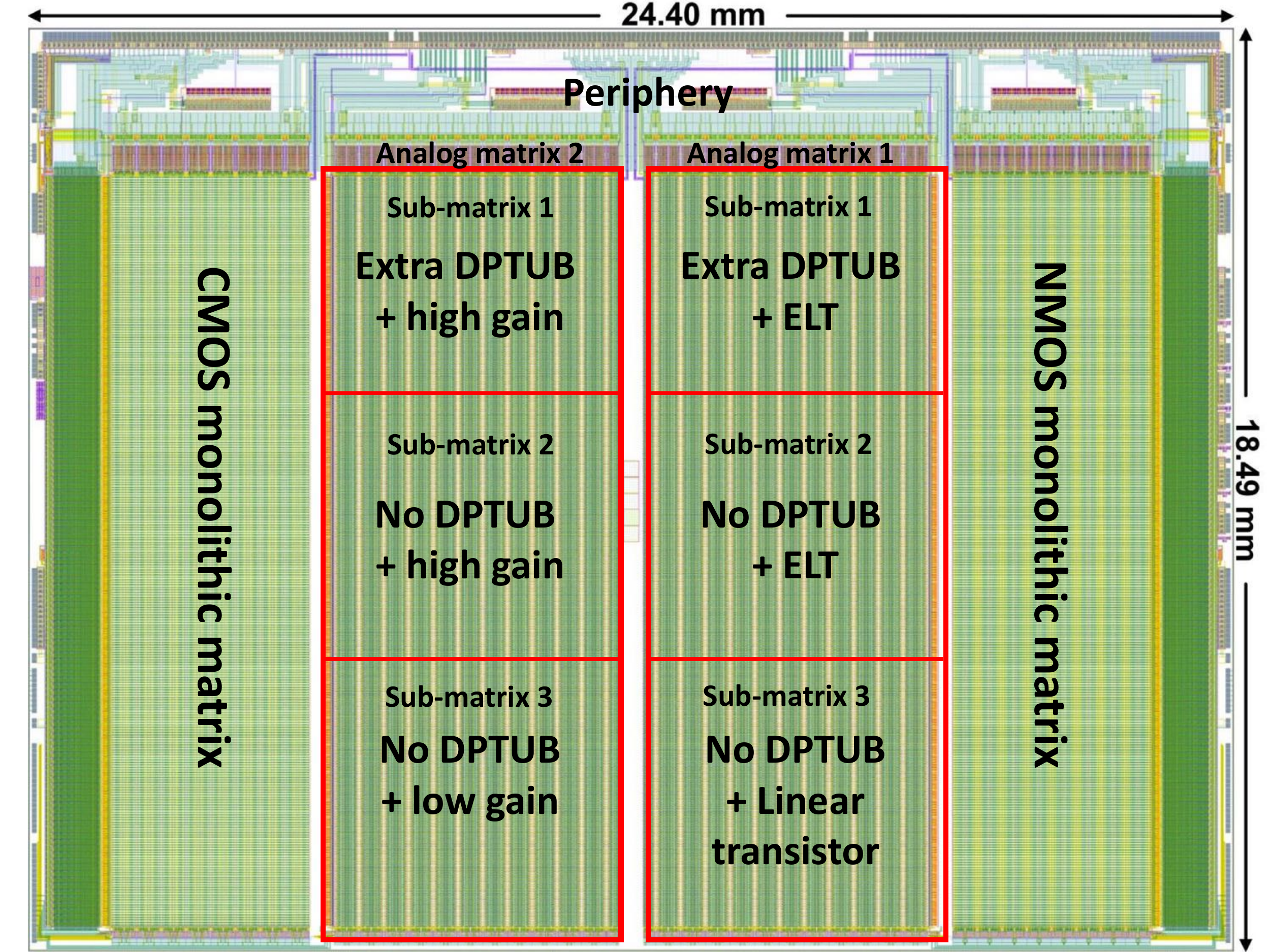}
  \caption{Floor plan of the H35DEMO demonstrator sensor. The monolithic matrices including sensor elements and readout electronics are located at the left and right. The two capacitively coupled detector matrices are located in the middle.}
  \label{H35DEMOLayout}
\end{figure}

The H35DEMO demonstrator chip \cite{H35DEMODesign} is a large size pixel sensor chip designed and produced in the ams H35 HV-CMOS technology using three types of high-resistivity substrates: \SIlist{80;200;1000}{\ohm\cm}. The design, as seen in figure \ref{H35DEMOLayout}, includes four independent sub-matrices: the NMOS and CMOS monolithic matrices integrating sensor and readout electronics into the same die and the two analog matrices 1 and 2 designed for capacitive-coupling to the FE-I4 readout ASIC \cite{FE-I4}, in order to decouple readout electronics aspects from sensor diode properties. The monolithic matrices were tested before and after irradiation and demonstrated good performance of the readout circuitry required for the monolithic integration of CMOS pixel sensors in ams HV-CMOS technology \cite{Terzo2017,Casanova2017}. The pixels in the analog matrix, measuring \SI{250x50}{\um}, contain a large collection diode in which the amplification circuitry is implemented. The amplified signal is then routed to an output pad matching the input of the FE-I4 ASIC in order to be detected and digitised. The large signal excursion of the amplifier allows for an efficient transmission of the signal through a small capacitor formed by a thin layer of glue between the H35DEMO and the FE-I4 pads. To achieve a good uniformity of the coupling between pixels over the whole matrix, a glueing process using high-precision flip-chip machine was developed and used.

\subsection{Analog pixel flavours}

\begin{figure}[!htbp]
  \center
  \includegraphics[width=0.685\textwidth]{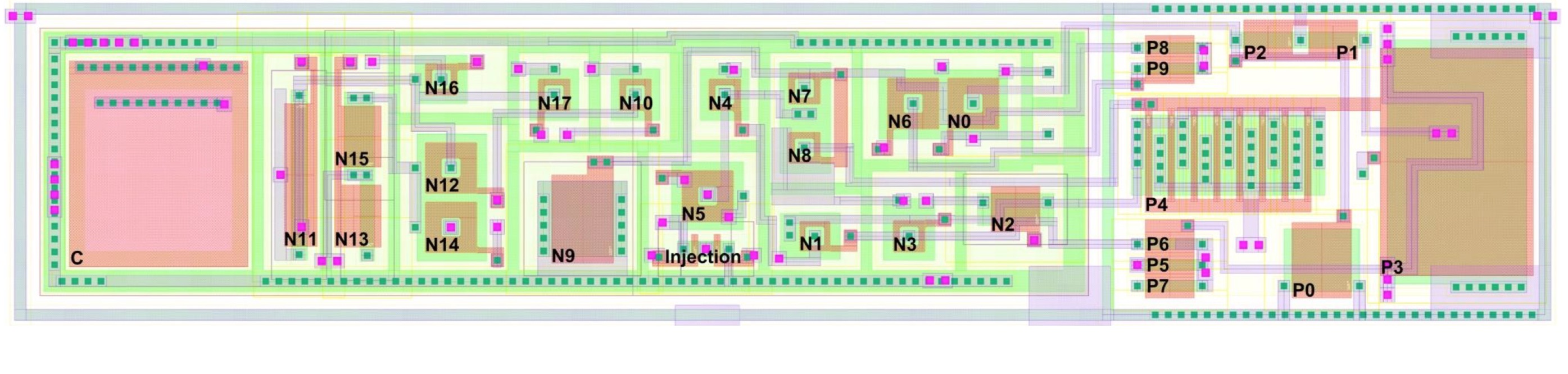}
  \includegraphics[width=0.67\textwidth]{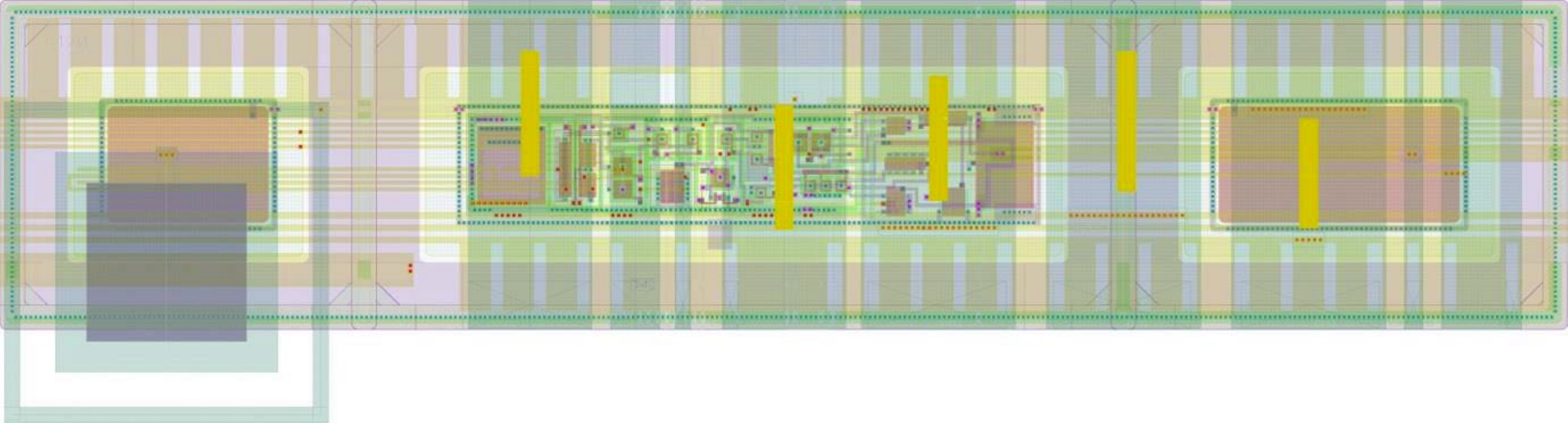}
  \includegraphics[width=0.7\textwidth]{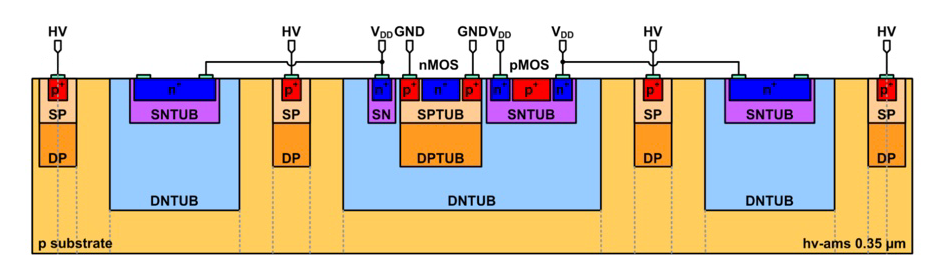}
  \caption{Top: Transistor layout (\SI{20.25x95.05}{\um^2}) of one flavour of the H35DEMO pixel in the Analog Matrix.  Middle: Amplifier and output stage transistors embedded in the central deep N-Well while to additional peripheral deep N-Well are added to each side of the central well to fill the pixel area with the collection electrode. The three deep N-Wells are electrically connected through metal to the input of the amplifier. The output top metal pad (dark blue square) is located on the extremity of the pixel with a small offset in the vertical direction. Bottom: doping structure of the H35DEMO analog pixel, illustrating the different doping structures. N-type implants are represented by the letter "n", while p-type is represented by "p". Shallow doping is indicated by "S" while deep implants are represented by "D". It is also possible to see the HV, GND and V$_{dd}$ input connections.}
  \label{H35DEMOpixel}
\end{figure}

The H35DEMO analog matrices were each sub-divided along the columns into three flavours of in-pixel amplification circuit of equal size (23x100 pixels). The first analog matrix consists of circuits containing different flavours of transistors, linear or enclosed layout (ELT) and different P-Wells (with or without the Deep P-Well (DP) illustrated  in figure \ref{H35DEMOpixel}) for high voltage biasing. The second analog matrix also contains three types of pixel, with different gain in the second stage of amplification and different P-Wells for high-voltage biasing as in the first matrix. The different flavours are summarised in table \ref{submatrices}. The H35DEMO prototype pixel amplifiers have not been optimised to produce a fast signal meeting LHC requirements, as the 3.3V power supply and the large capacitance of the pixels matching the FE-I4 pixel size would yield to a large power consumption complicating the operation of the sensor. 

\begin{table}
\centering
\caption{Sub-matrix flavours in H35DEMO analog sensors}
\label{submatrices}
\begin{tabular}{|p{2cm}|p{5.5cm}|p{5.5cm}|}
\toprule
  & Matrix Analog 1 (ANA1) & Matrix  Analog 2 (ANA2) \\
\midrule
sub-matrix 1& - Extra DPTUB for High Voltage \newline - ELT in the feedback circuitry & - Extra DPTUB for High Voltage \newline - High gain \\ \hline
sub-matrix 2& - Without DPTUB for high voltage \newline - ELT in the feedback circuitry & - Without DPTUB for high voltage \newline - High gain \\\hline
sub-matrix 3 & - Without DPTUB for high voltage \newline - Linear transistors in feedback & - Without DPTUB  for high voltage \newline - Low gain\\
\bottomrule
\end{tabular}
\end{table}

\subsection{Capacitive coupling to the FE-I4 ASIC}
\begin{figure}[!htbp]
\center
  \includegraphics[width=0.95\textwidth]{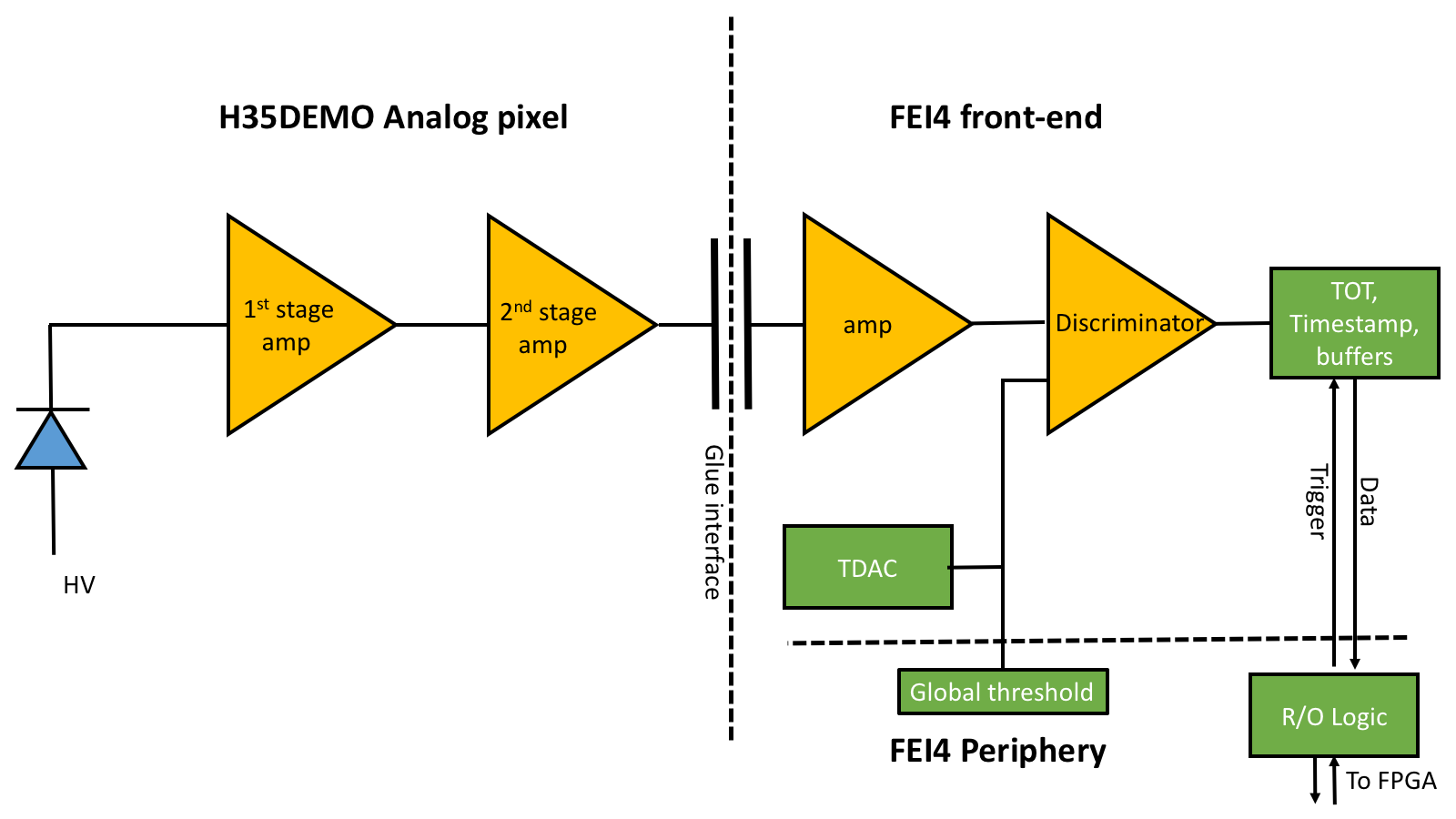}
  \caption{Schematic representation of the readout chain of the H35DEMO prototype capacitively coupled to FE-I4.}
  \label{H35DEMOpixelScheme}
\end{figure}

In capacitively coupled hybrid detectors, the signal processed in the sensor is transmitted to the read-out chip (ROC) via a capacitive injection. The pixel pads at the top surface (last metal layer) of the sensor and of the ROC are aligned with each other and coupled together by a thin layer (from \SI{0.2}{\um} up to \SI{10}{\um}, depending on the bonding force) of a non-conductive resin by the flip-chip process. Each pixel pad will act as a capacitor terminal with the glue as a dielectric layer in between. The particle signal in the sensor is transmitted as a voltage pulse generated in the pixel pad that  then creates a charge pulse signal in the ROC pixel pad, where it is followed by a charge sensitive amplifier and so on, as illustrated in figure \ref{H35DEMOpixelScheme}.

\begin{figure}
\begin{minipage}[b]{0.51\textwidth}
\begin{center}
\includegraphics[width=1\textwidth]{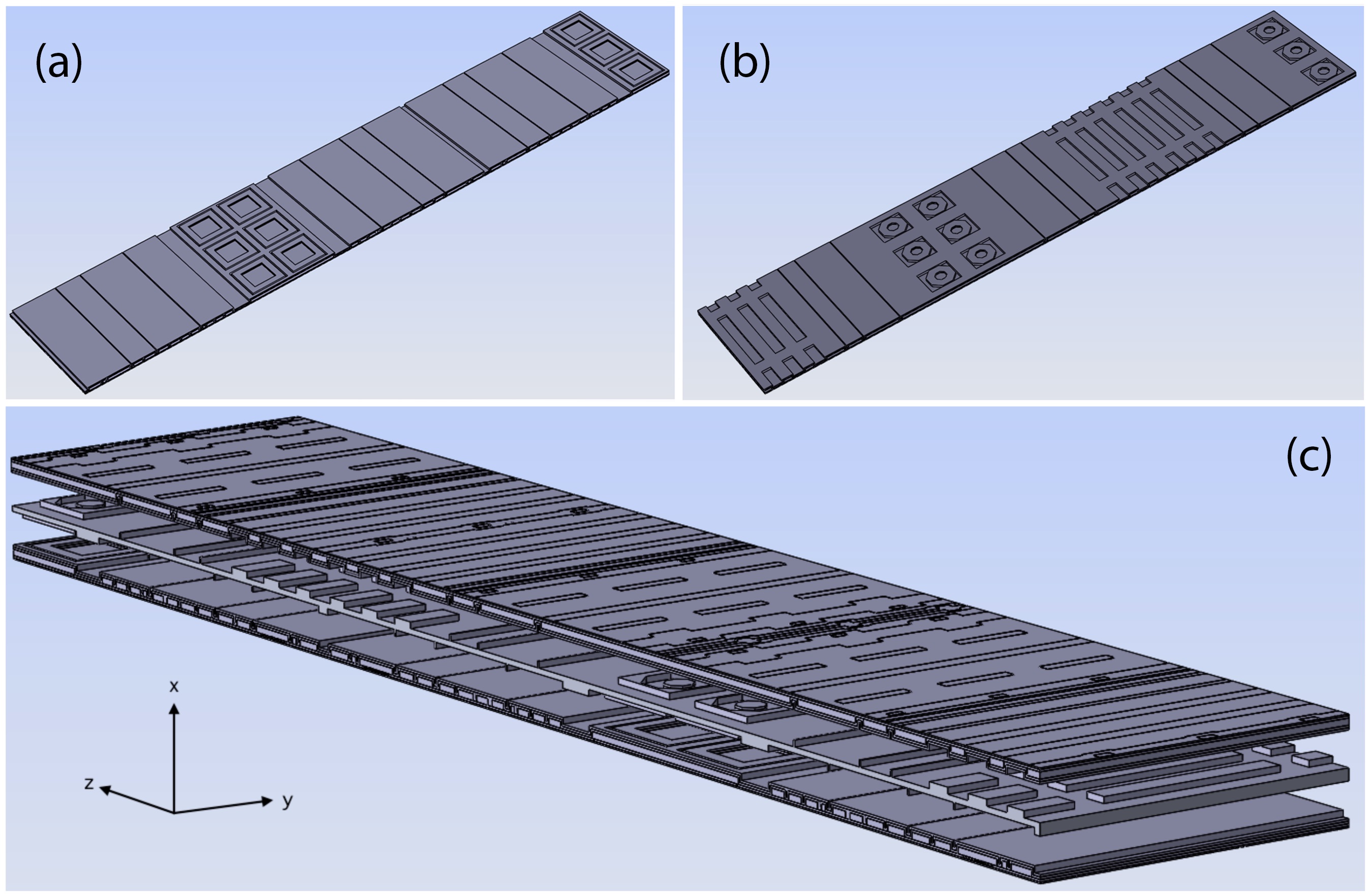}
\end{center}
\caption{(a) H35DEMO 3x3 pixel matrix. (b) FE-I4 3x3 Pixel matrix. (c) 3x3 matrices facing each other, with the conformal dielectric glue layer in between.}
\label{3dgeo}
\end{minipage}
\hspace*{\fill} % it's important not to leave blank lines before and after this command
\begin{minipage}[b]{0.43\textwidth}
\begin{center}
\includegraphics[width=1\textwidth]{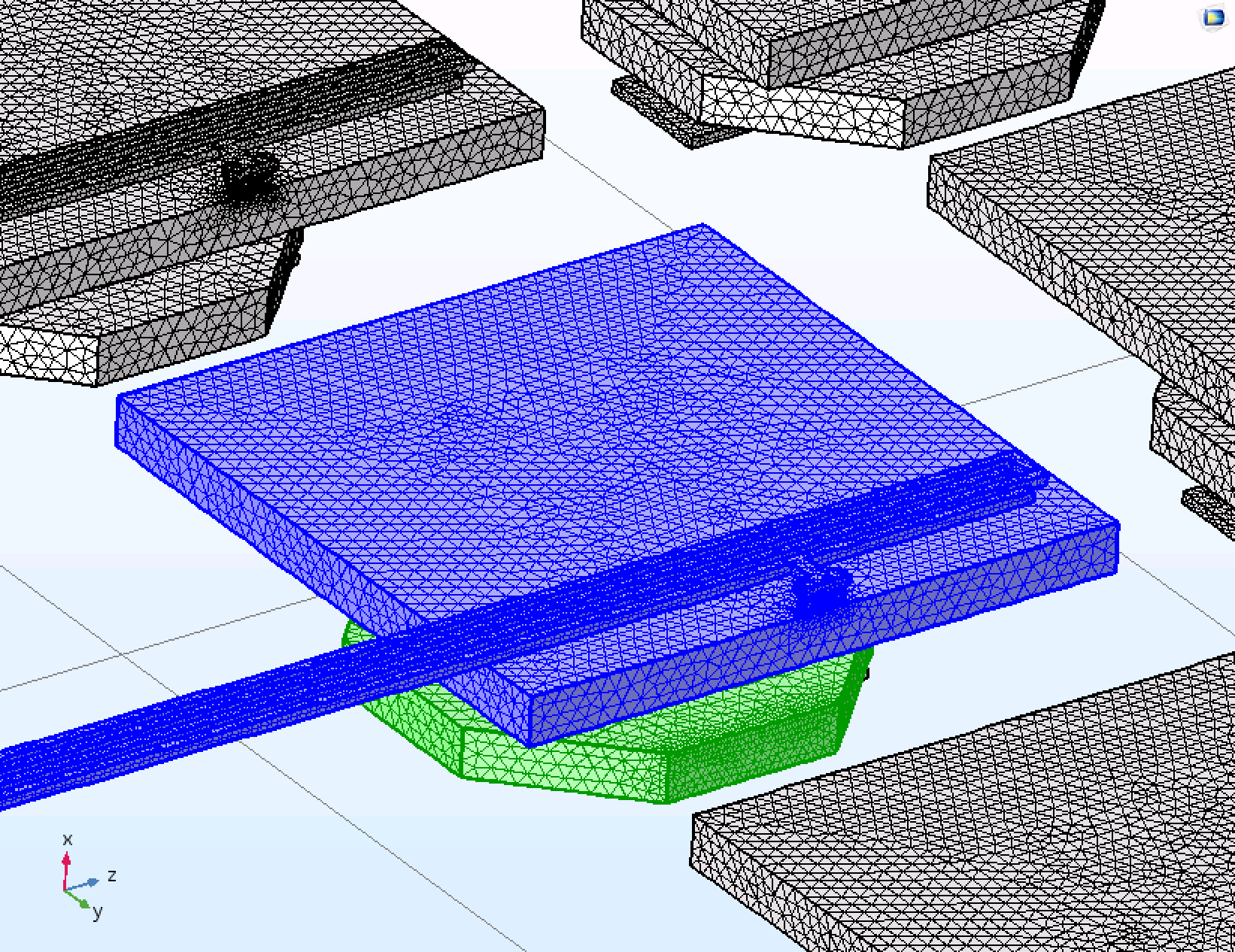}
\end{center}
\caption{H35DEMO pixel pad (blue) and FE-I4 pad (green), plus neighbouring pixels, simulation mesh.}
\label{pixel_pads}
\end{minipage}
\end{figure}

As the method of signal transfer between the sensor and the ROC is via a capacitive injection, potential crosstalk to neighbouring pixels must be considered. For that, a 3D simulation of the coupling between the pixel pads is performed with the COMSOL Multiphysics software, using the Finite Element Analysis method. A detailed 3D 3x3 pixel matrix was modeled based on each chip GDSII design file. The model includes the CMOS stack from the 3rd metal layer up to the last passivation layer. Figure \ref{3dgeo} shows the 3D geometry created for the simulation. 

\begin{table}[ht]
  \begin{varwidth}[b]{0.45\linewidth}
    \centering
    \begin{tabular}{ c r r r r}
      \toprule
FE-I4 Pixel & Capacitance [fF] & Cross-talk [\%]\\    \midrule
1 & $10^{-63}$ & $10^{-61}$ \\
2 & $10^{-62}$ & $10^{-60}$ \\
3 & $10^{-62}$ & $10^{-61}$ \\
4 & $5.5 \times 10^{-4}$ & $1.60\times 10^{-2}$ \\
5 & 3.46 & 100 \\
6 & $5.54 \times 10^{-4}$ & $1.60\times 10^{-2}$ \\
7 & $6.42\times 10^{-7}$ & $1.86\times 10^{-5}$ \\
8 & $5.61\times 10^{-4}$ & $1.62\times 10^{-2}$ \\
9 & $6.68\times 10^{-7}$& $1.93\times 10^{-5}$ \\
      \bottomrule
    \end{tabular}
    \caption{Coupling capacitances between 9 FE-I4 pixels and a H35DEMO pixel with a 2.5 $\mu$m gap in between.}
    \label{cap_matrix}
  \end{varwidth}%
  \hfill
  \hfill
  \hfill
  \hfill
  \hfill
  \hfill
  \hfill
  \begin{minipage}[b]{0.45\linewidth}
    \centering
    \includegraphics[width=1.1\textwidth, trim= 0 0 0cm 0cm]{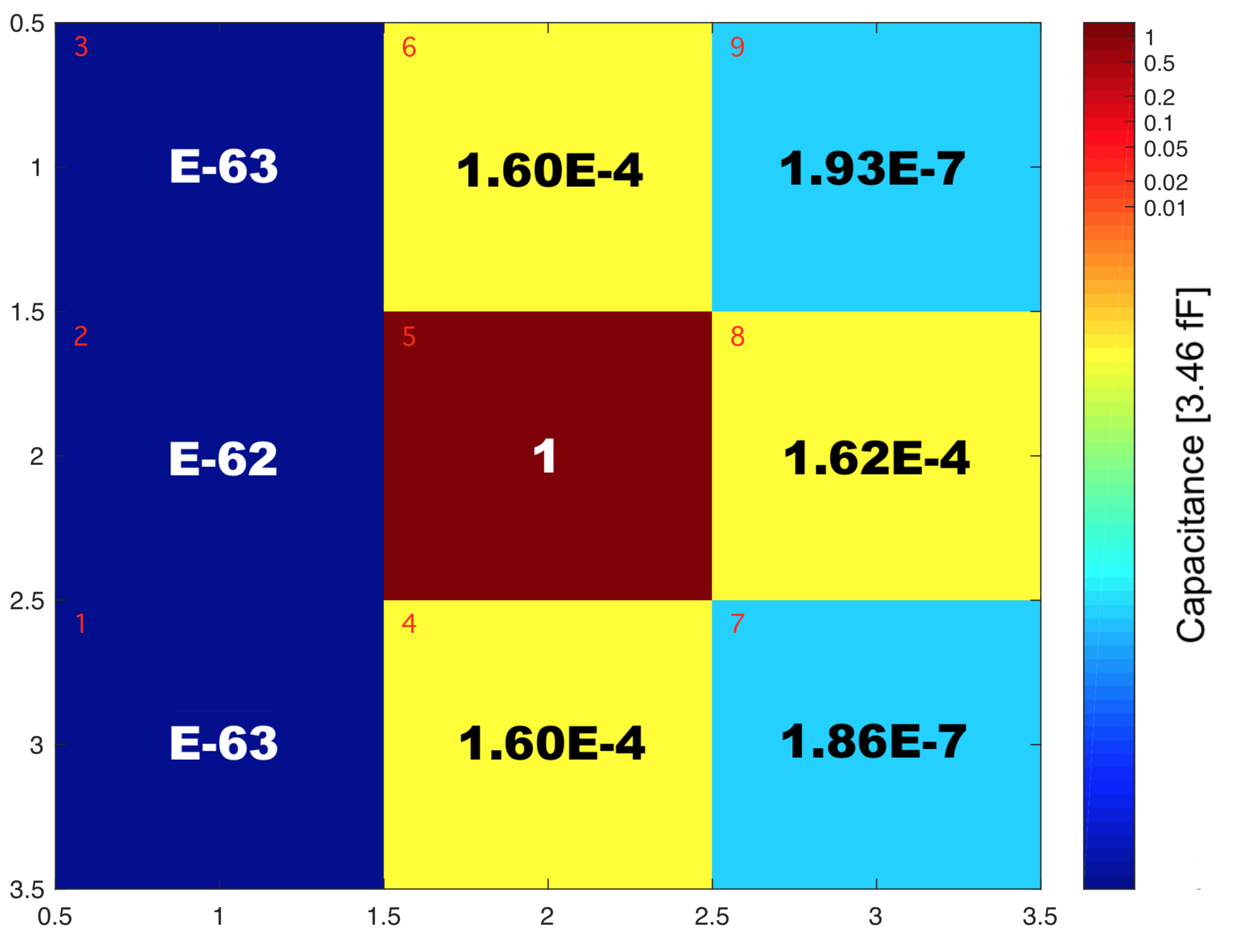}
    \captionof{figure}{Normalized coupling between H35DEMO pixel and 9 FE-I4 pixels.}
    \label{cap_map}
  \end{minipage}
\end{table}

COMSOL simulates the electric field between the pixel pads, shown on figure \ref{pixel_pads}, and calculates the capacitance between all pads. Table \ref{cap_matrix} lists the coupling capacitances between the center pixel of the 3x3 H35DEMO matrix with all the 9 pixels in the 3x3 FE-I4 pixel matrix. As the charge transferred between the sensor and the ROC is linearly proportional to the capacitance, the 3rd column of table \ref{cap_matrix} shows the coupling capacitance normalized with respect to the middle pixel of the FE-I4 matrix, and the same relative coupling is also shown on figure \ref{cap_map}, helping to visualize the amount of charge that would be transferred due to the cross-coupling to neighbouring pixels. The asymmetry observed in coupling capacitance is due to the layout of pixels in the FE-I4 in double columns , reproduced in the model, as can be seen in figure \ref{3dgeo} a) and b). This geometry allow to extract the coupling capacitance for the pixels inside a double column and between double columns. The results from the simulation shows that the expected coupling capacitance for the main pixel is in the order of 3.5 fF, with a maximum cross-coupling around 0.016\%, meaning that the charge induced to the neighbouring pixels, due to cross-talk, will mostly be under the detection threshold.

\begin{figure}[!htbp]
  \subfigure{
  \includegraphics[width=0.29\textwidth]{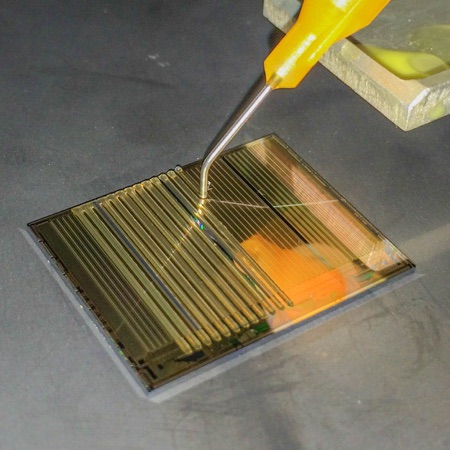}
  }
  \subfigure{
  \includegraphics[width=0.29\textwidth]{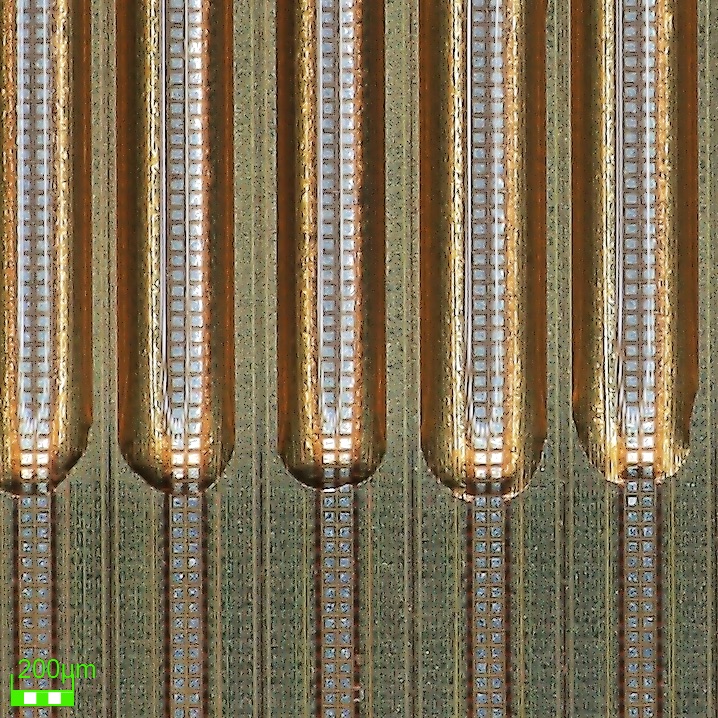}
  }  
    \subfigure{
  \includegraphics[width=0.38\textwidth]{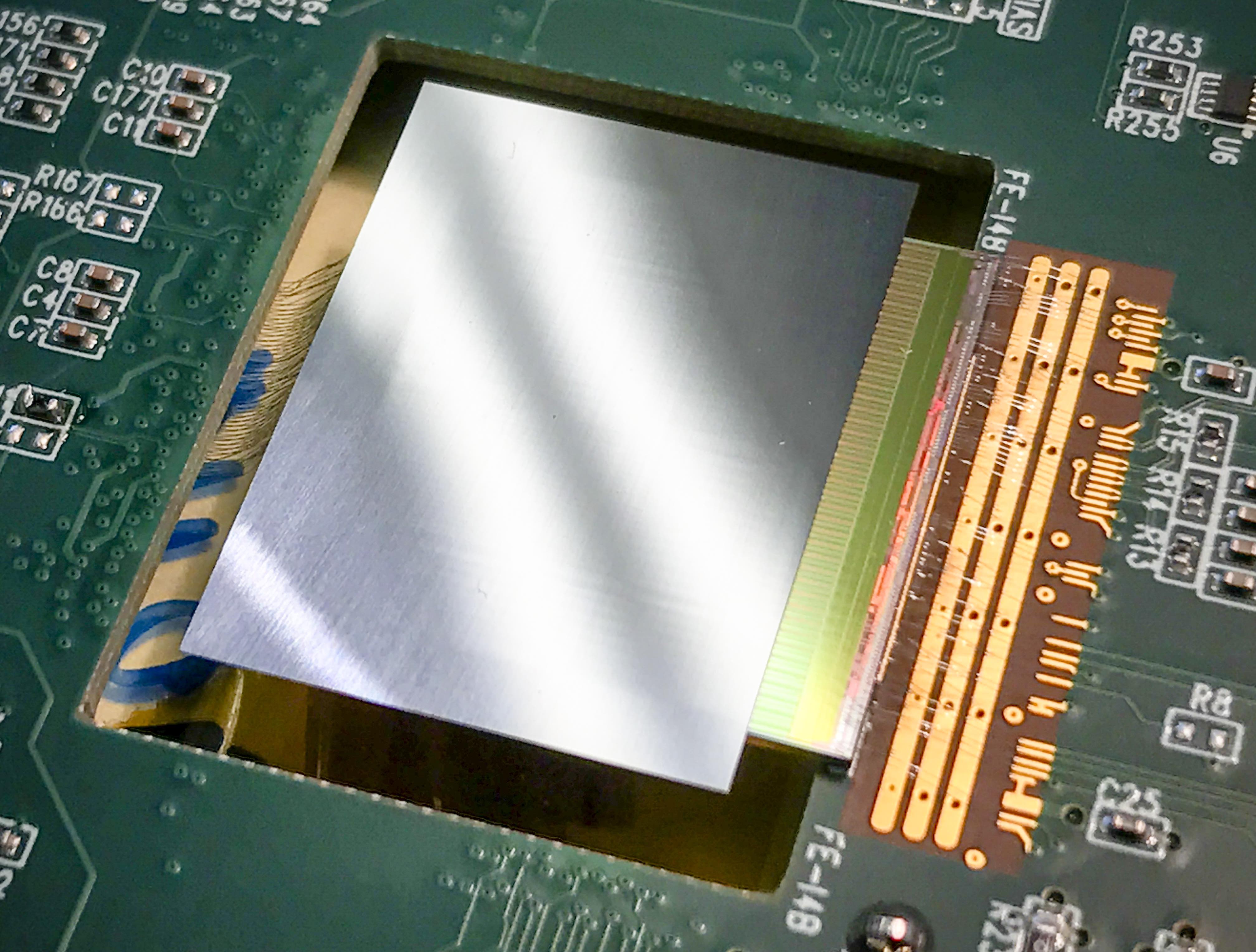}
  }  
  \caption{Left: deposition of epoxy on the H35DEMO matrix by the
    automatic glue time-pressure dispenser of the Acc$\mu$ra 100. Middle: glue (partially) deposited on double pixel column. Right: 100 ~\micron thin H35DEMO-FE-I4 assembly on PCB.}
  \label{glue}
\end{figure}

The glueing process was developed using the $\mathrm{Acc\mu ra~100}$ flip-chip bonder. The machine controls the bonding with a precision of \SI{+-1.5}{\um} and a parallelism of \SI{+-1}{\micro\radian}. Araldite 2011 epoxy was used as the adhesive and coupling medium. Forty lines of glue, one for each FE-I4 double-column, were dispensed at a speed of \SI{2}{\cm\per\s} using a \SI{3}{cc} syringe with a \SI{600}{\um} dispensing tip and \SI{4}{\bar} of pressure in the time-pressure dispenser. Figure \ref{glue} shows the dispensing process inside the flip-chip machine and the final assembly mounted on PCB.

\begin{figure}
\begin{minipage}[b]{0.42\textwidth}
\begin{center}
  \subfigure{
  \includegraphics[width=0.95\textwidth]{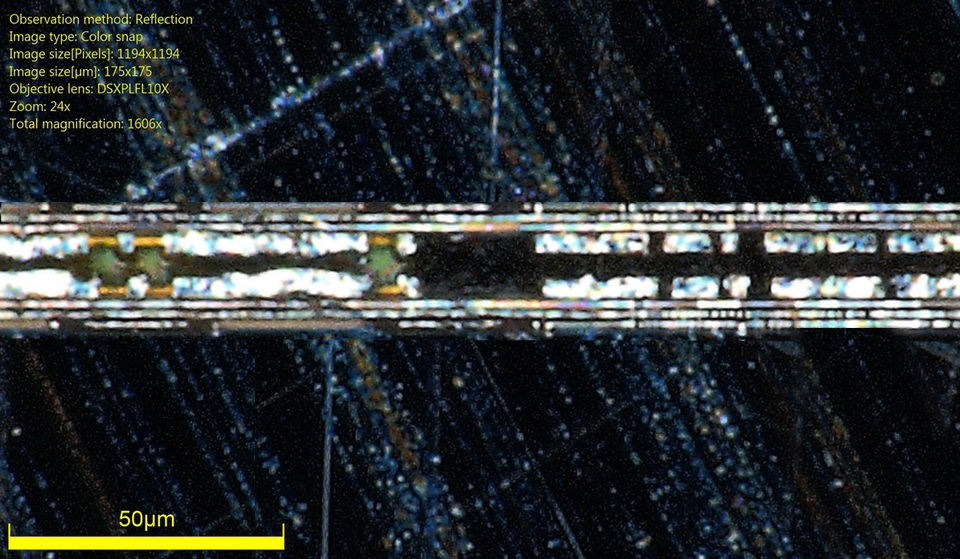}
  }  
  \subfigure{
  \includegraphics[width=0.95\textwidth]{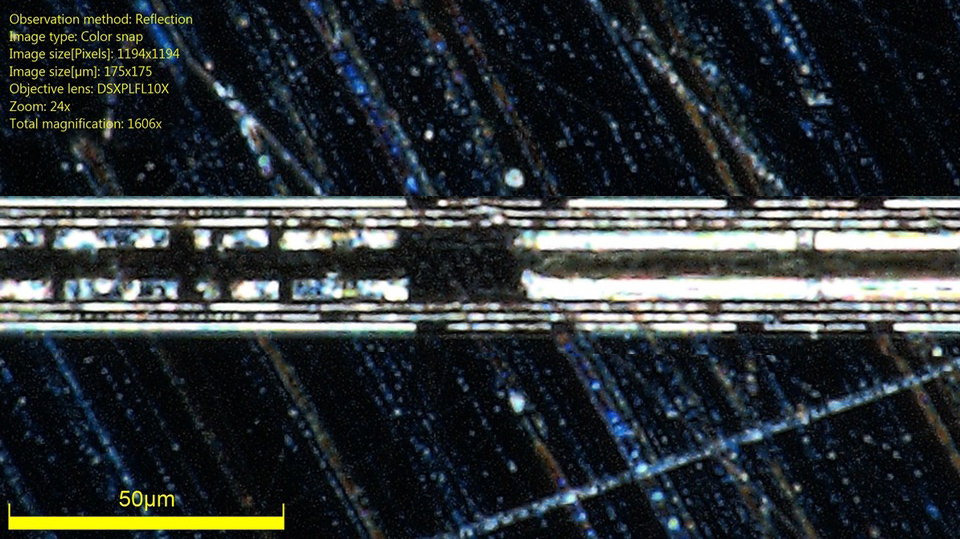}
  }\end{center}
\caption{Thickness of the glue layer along the chip edge at two locations along the chip at \SI{2}{\cm} distance  showing good parallelism, less than \SI{100}{\nm} difference measured with an optical microscope, from left to right.}
\label{crosssection}
\end{minipage}
\hspace*{\fill} % it's important not to leave blank lines before and after this command
\begin{minipage}[b]{0.53\textwidth}
\begin{center}
\includegraphics[width=0.95\textwidth]{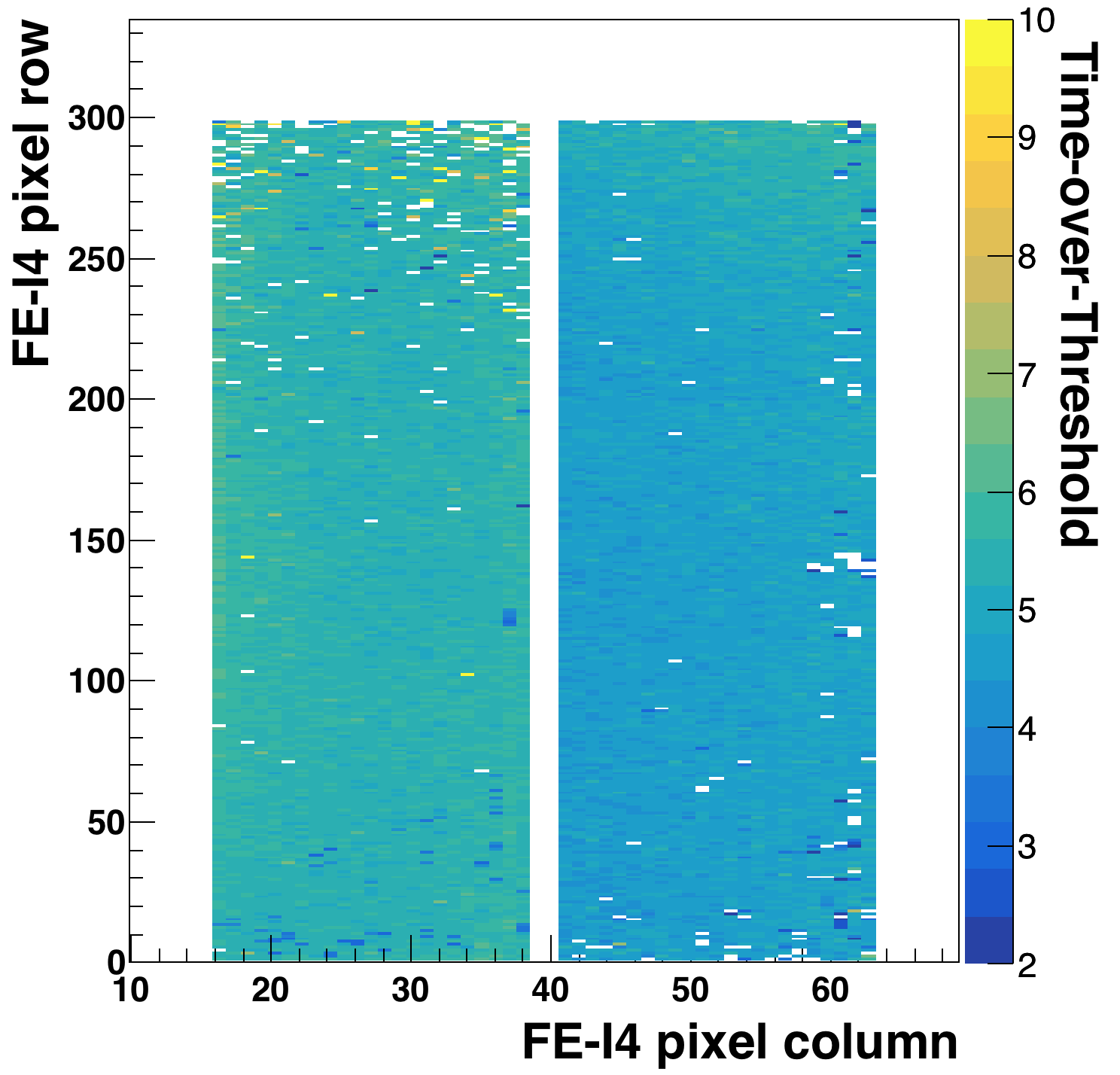}
\end{center}
\caption{Combined ToT map from analog matrix 1 (left) and 2 (right). The lower threshold on the second analog matrix is due to the higher FE-I4 threshold (\SI{3000}{e}), comparing with the \SI{2000}{e}threshold for the first analog matrix, during data acquisition.}
\label{tot_map}
\end{minipage}
\end{figure}

During the production, several mechanical samples were assembled and cross-section and metrology studies were performed on the glue interface to
verify the parallelism and measure the distance between the H35DEMO and FE-I4 pads. Figure \ref{crosssection} shows the cross-section image for one of these samples on both sides of the assembly, at \SI{2}{\cm} distance. The measurement demonstrated that the process we developed produces assemblies with a good parallelism, i.e. less than \SI{100}{\nm} difference in silicon to silicon distance over the whole assembly, and good control of the distance between the coupling pads with the passivation of each pad in contact with its respective pad on the FE-I4 ASIC. 

Later, data taken during testbeam also confirmed the good flip-chip parallelism as shown on the ToT map on figure \ref{tot_map}. The amplitude coming from the pixel output has been simulated to be on the range between 100 mV, for an injected charge of \SI{750}{e} (0.5 MIP), and 300 mV for \SI{4500}{e} (3 MIP) \cite{H35DEMODesign}, with a noise corresponding to \SI{120}{e}. With the simulated H35DEMO pad coupling capacitance of \SI{3.5}{\fF}, the current signal generated in the FE-I4 amplifier (by a MIP particle) will be in the order of \SI{2000}{e}. The FE-I4 amplifier was tuned to yield a ToT = 10 for an input charge of \SI{16}{e}. Figure \ref{tot_map} combines the data taken separately from analog matrix 1 and 2 and shows that an uniform ToT distribution is achieved on both analog matrices. The lower ToT on the second analog matrix is due to the higher threshold used during data acquisition with ANA2, where the FE-I4 threshold was set to \SI{3000}{e}, while the FE-I4 threshold was set at \SI{2000}{e} for ANA1.

\section{Test beam experimental setup and reconstruction}

The Geneva FE-I4 Telescope \cite{telescopePaper} was used for the
measurements of the H35DEMO prototypes. It comprises six telescope
planes built from ATLAS IBL hybrid sensors, i.e. a planar passive
silicon sensor bump-bonded to a FE-I4 readout ASIC. It is placed inside
a particle beam and the six planes are used to measure the beam
particles independent from the device-under-test (DUT). A RCE readout
system \cite{rce} was used for the data acquisition of the telescope
planes and the DUT. The hit bus signal of the first and the last
telescope planes are combined to provide a trigger for the data acquisition
system. The CaRIBOu system \cite{Caribou} was used for the slow control
of the H35DEMO prototypes. DUT samples were mounted inside a thermally
insulated box equipped with cooling provided by a chiller and maintained
at \SI{25}{\celsius} or less during data taking.

The reconstruction and analysis of the data obtained were performed
using the Proteus reconstruction software \cite{Proteus}. Proteus starts
with the raw hit data and provides fully reconstructed clusters and
tracks. First, it combines neighboring hits on each sensor into clusters
using a greedy clustering algorithm. Then, it estimates the cluster
position and cluster signal. The specific algorithms can be configured
for each sensor type and here both the telescope sensors and the DUT use
the time-over-threshold-weighted center-of-gravity as a position
estimator.

For track finding, clusters on the telescope planes are transformed into
the global coordinate system using a three-dimensional geometry
description of all planes that takes into account all degrees of freedom
of a planar surface in three-dimensional space. Starting from clusters
on the first sensor, track candidates are found by extrapolating the
initial position to all further telescope planes along the beam
direction and adding matching clusters. Ambiguities are solved by
bifurcation of the track candidate. From all candidates, tracks are
selected by exclusively associating clusters to tracks starting from the
longest track with the lowest fit $\chi^2$ value.

All planes are aligned first using a rough alignment based on
correlations and then using a track-based alignment that minimises track
residuals. To provide a consistent performance over long data taking
periods, the alignment procedure is performed for each run. The first
\num{20000} events are used to align the geometry and are not used
during the analysis.

For the analysis, only tracks with clusters on all six telescopes and a
$\chi^2/d.o.f.$ below \num{5} are considered. All further operations are
performed in a local coordinate system anchored on the surface of the
DUT. Tracks are reconstructed by performing a weighted least squares fit
in the local coordinate system using a linear track model. The
reconstructed positions on the DUT surface have a resolution of
\SI{10}{\um} along the column direction and \SI{12}{\um} along the
row direction of the H35DEMO prototype. Reconstructed tracks are then
matched to H35DEMO clusters using a matching cut on the distance between
the two of $\SI{250}{\um}$ along each axis.

Samples of three different resistivities (\SIlist{80;200;1000}{\ohm\cm}) were studied with bias voltage scanned from \SIrange{0}{160}{\V} and FE-I4 thresholds from \SIrange{1000}{4000}{e}. For each point, one to ten million triggers were recorded.

\begin{table}
\center
\begin{tabular}{|c|c|c|l|}
\toprule
Resistivity (\si{\ohm\cm}) & Matrix & Bias Voltages (\si{\V}) & FE-I4 Threshold (\si{e})\\
\midrule
80 & Analog 1 & 0-160 &1500, 2000  \\ \hline
200 & Analog 1 & 0-160 & \hspace*{1.2cm} \hspace*{0.5cm} 2500, 3000, 4000 \\ \hline
200 & Analog 2 & 0-140 &\hspace*{0.85cm} 2000, 2500, 3000, 4000 \\ \hline
1000 & Analog 1 & 0-160 & 1500, 2000, 2500, 3000 \\
\bottomrule
\end{tabular}
\caption{Summary of measurements performed on the H35DEMO prototypes in
  test beam. The Analog 2 matrix of the \SI{1000}{\ohm\cm} sample
  was not measured due to an early increase in leakage current below the
  breakdown voltage as described later in the text. Note that the FE-I4 threshold regards the signal being transferred by the capacitive injection from the H35DEMO, instead of the real charge generated on the sensor.}
\end{table}

\section{Results}

\subsection{Current-Bias (I-V) characterisation}

\begin{figure}[!htbp]
\center{
\vspace*{-0.8cm}
  \subfigure[t][\SI{80}{\ohm\cm} substrate]{\includegraphics[width=0.49\textwidth]{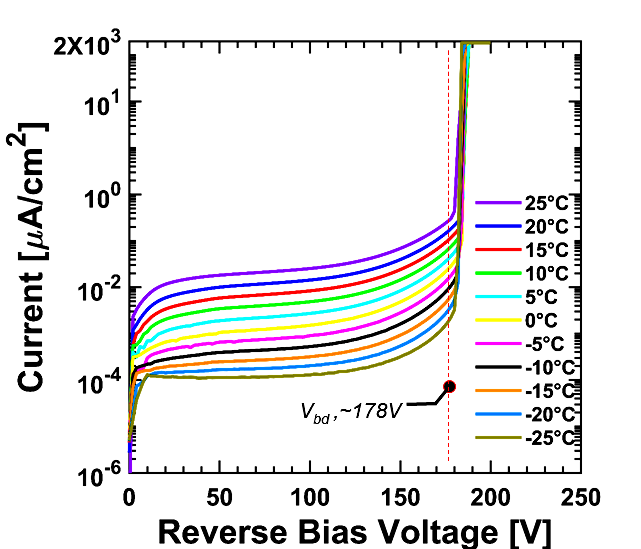}}
  \subfigure[t][\SI{200}{\ohm\cm} substrate]{\includegraphics[width=0.49\textwidth]{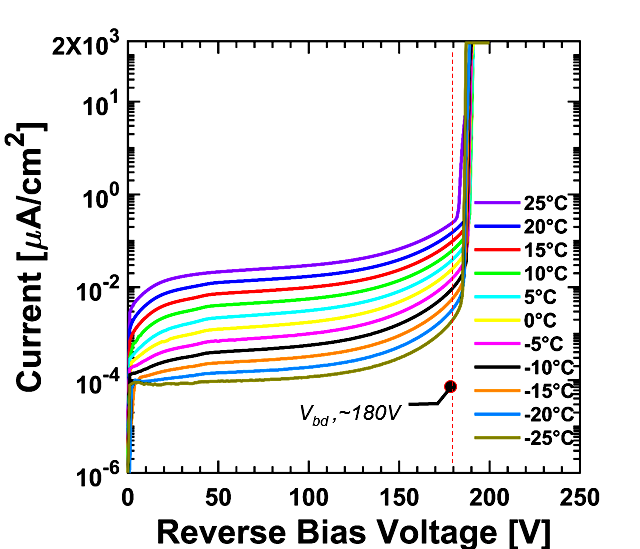}}
  \newline
  \subfigure[t][\SI{1000}{ \ohm\cm} substrate]{\includegraphics[width=0.49\textwidth]{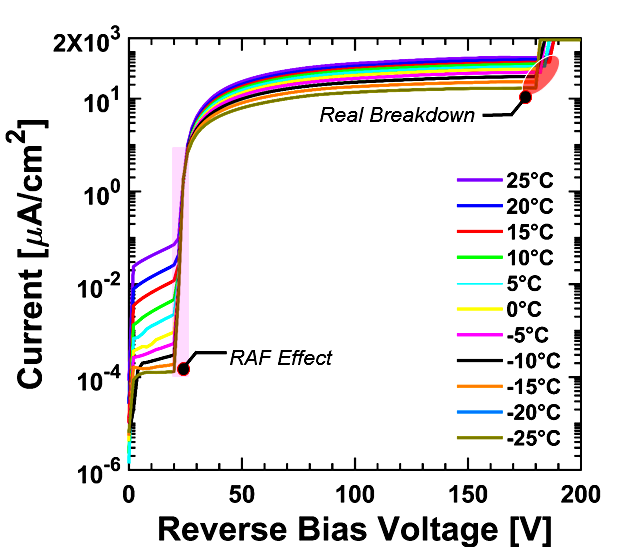}}
  \subfigure[t][Arrhenius plot for \SI{1000}{ \ohm\cm} substrate]{\includegraphics[width=0.49\textwidth]{plots/Arrhenius}}}
\vspace*{-0.3cm}\caption{Characteristic I-V curves and Arrhenius plot for \SIlist{80;200;1000}{\ohm\cm} substrate resistivity samples for temperatures
  ranging from \SIrange{-25}{25}{\celsius}. Leakage current dependence on temperature as a function of the inverse of the temperature for a \SI{1000}{\ohm\cm} substrate, also known as the Arrhenius plot, is presented in figure d). Leakage current differs from the prediction from the Shockley-Read-Hall (SRH) model of Generation-recombination of thermal carriers in a silicon bulk \cite{srh}.}
\label{IVs}
\end{figure}

The H35DEMO samples tested were first characterised using a probe station to verify the quality of the assemblies and  measure the leakage current versus device temperature when high voltage was applied to the sensor. Figure \ref{IVs} shows the I-V curves for the different resistivities under study. The \SI{80}{\ohm\cm} and \SI{200}{\ohm\cm} show low leakage current well below \SI{1}{\micro\A\per\cm\squared} for all temperatures up to the breakdown voltage, determined to be \SI{178}{\V} and \SI{180}{\V}, respectively. For the \SI{1000}{\ohm\cm} sample, an early offset of current is observed at a bias voltage of \SI{30}{\V}. This was initially mistaken for an early breakdown. However, further measurements have shown that a plateau of current is reached in the device before a second upset in current at \SI{180}{\V} is observed, corresponding to the real breakdown. This effect is known as the Rise-And-Flatten (RAF) effect \cite{RAF}. This additional leakage current corresponds to a surface current generated in other unbiased test structures of the H35 submission located at the periphery when reached by the depletion zone, larger laterally for the higher resistivity substrate. An Arrhenius plot was extracted from the data at different temperatures confirming that the generated leakage current does not correspond to a generation current from the bulk of the sensor. This effect could be limited in the future by removing or placing further the structures outside of the main matrix. As this phenomenon was discovered later during the testbeam campaign, due to time constraints only the first analog matrix of the \SI{1000}{\ohm\cm} was studied with a bias voltage larger than to \SI{30}{\V}, the second matrix show same behaviour but was not studied due to limited beam time.

\begin{figure}[!htbp]
\vspace*{-0.1cm}
  \subfigure[\SI{200}{\ohm\cm} Analog Matrix 1 efficiency $vs$ HV bias.]{
  \includegraphics[width=0.49\textwidth]{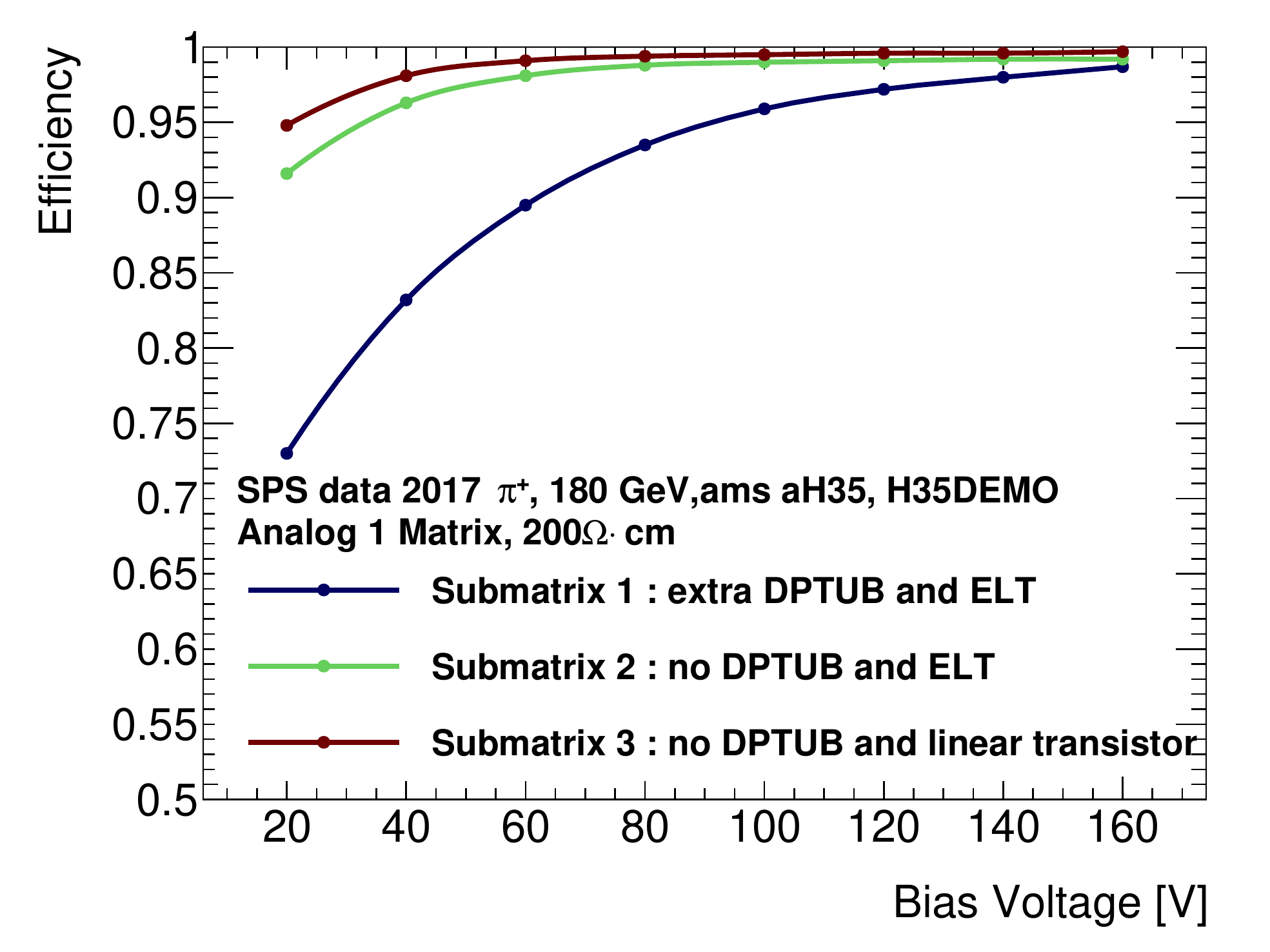}
  \label{ana1}}
  \subfigure[\SI{200}{\ohm\cm} Analog Matrix 2 efficiency $vs$ HV bias.]{
  \includegraphics[width=0.49\textwidth]{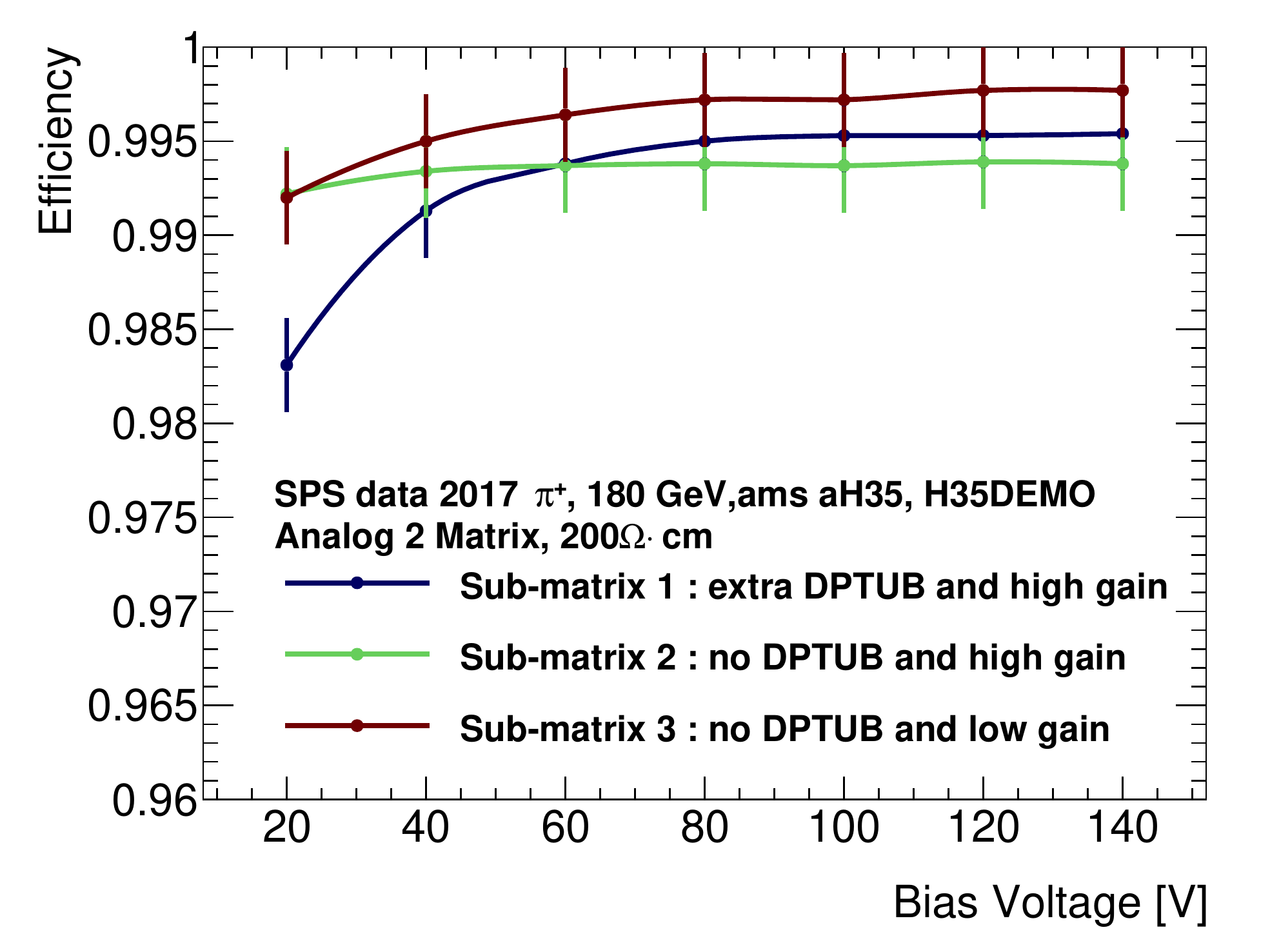}
  \label{ana2}}
  \vspace*{-0.3cm}\caption{Detection efficiency as a function of bias voltage for the three sub-matrices of analog matrix 1 (left) and analog matrix 2 (right) for a \SI{200}{\ohm\cm} sample for a threshold of \SI{2000}{e}.}
  \label{submatrix_eff}
\end{figure}

\subsection{Comparison of pixel flavours}

The H35DEMO analog matrices contain six flavours of pixels with different
gain, feedback and biasing schemes. The purpose of these variations is
to determine the best scheme to ensure a low noise and a high-efficiency
operation of the pixel. The Deep P-Well, a deep implant located below
the high voltage implant (see figure \ref{H35DEMOpixel}) influences the
pixel input capacitance and therefore the noise and rise time of the
signal. Enclosed layout transistors (ELT) are used to make the circuitry more
radiation tolerant to ionising dose but will negatively affect the gain
and rise-time of the pulse generated by the circuitry. The gain, as
determined by the feedback capacitor in the pixel amplifier, will affect
the noise and the detection efficiency, for a given injection
charge. Figure \ref{submatrix_eff} shows the detection efficiency for
all sub-matrices of the analog matrices of the \SI{200}{\ohm\cm}
sample, as a function of the bias voltage applied to the sensor. The
detection efficiency is defined as the ratio between number of
reconstructed tracks matched to a cluster on the DUT and the total
number of reconstructed tracks in the acceptance. Figure \ref{ana1} shows that the use of an ELT in the feedback circuitry of the pixel (sub-matrix 2 )doesn't impact significantly the pixel detection efficiency when compared with the pixel flavour using linear transistors (sub-matrix 3). The most significant effect is however linked to the addition of the Deep P-Well implant to the high-voltage implant, degrading the detection efficiency in both analog matrices. The addition of capacitance between the Deep N-Well and the biasing contact due to this implant affects the gain and rise-time of the amplifier and should be avoided for particle detection. In Analog matrix 2, for the three cases, efficiency is well above \SI{99}{\%} in the conditions of figure \ref{ana2}.

Another important aspect that must be assessed with regard to the different pixel types is the timing accuracy as measured by the FE-I4 ASIC that relates to the rise time of the the pulse and the associated time walk due to the detection threshold.

\subsection{Cluster size, spatial resolution and time resolution}

\begin{figure}[!htbp]
\centering
\vspace*{-1cm}
  \subfigure[\SI{80}{\ohm\cm}]{%
\includegraphics[width=0.49\textwidth]{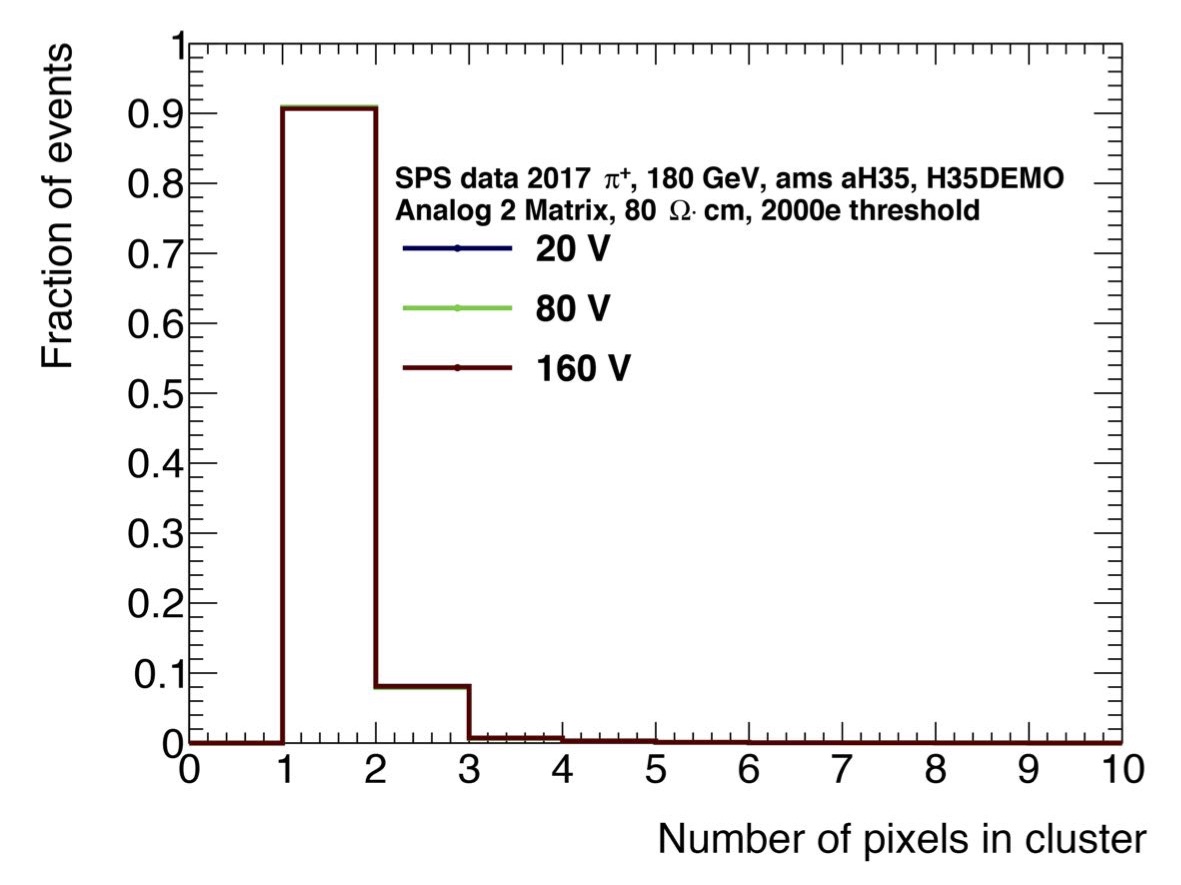}}
\hfill
  \subfigure[\SI{200}{\ohm\cm}]{%
\includegraphics[width=0.49\textwidth]{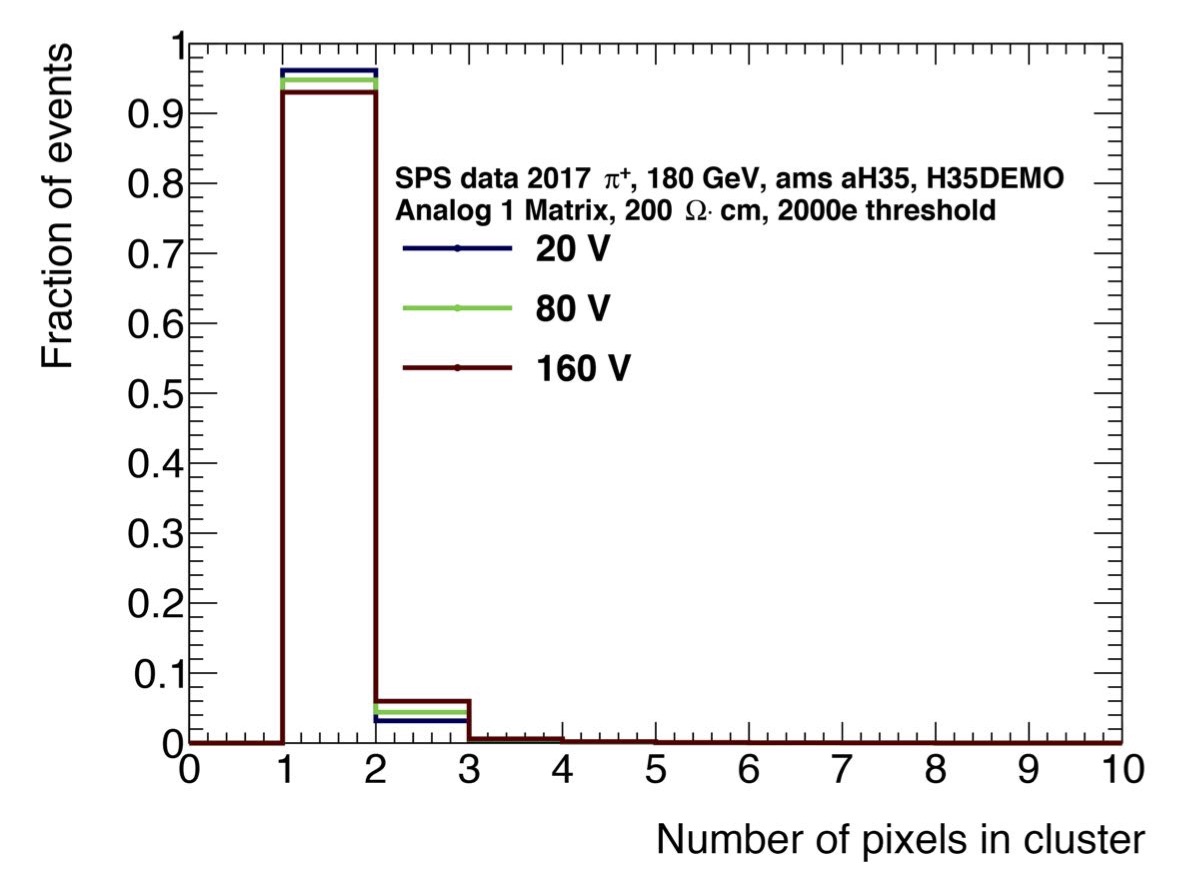}} \newline
\hfill     
  \subfigure[\SI{1000}{\ohm\cm}]{%
\includegraphics[width=0.49\textwidth]{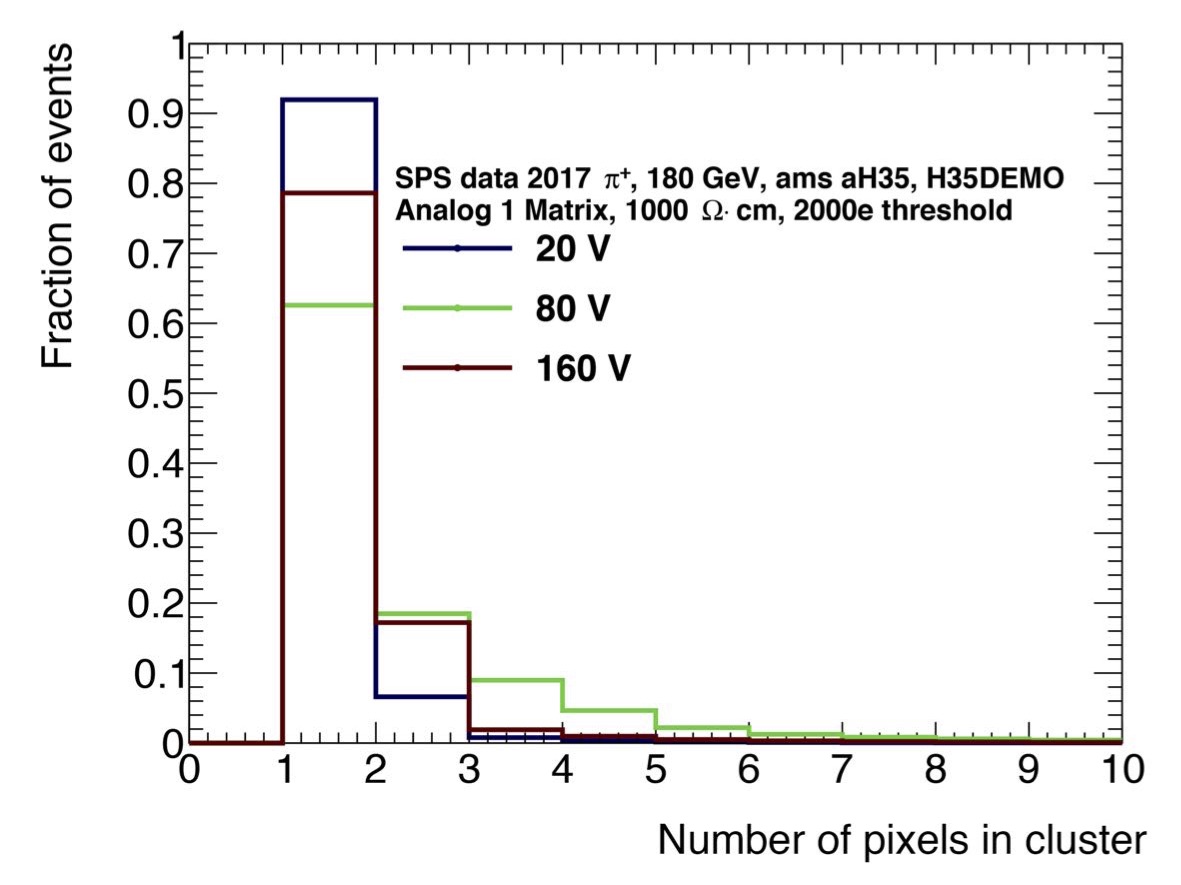}}\label{8c}
\caption{Cluster size versus bias voltage and different substrate
  resistivity for \SI{2000}{e} threshold.}
\label{CS_vs_resistivity}
\end{figure}

The effect of different substrate resistivity on the behaviour of the
H35DEMO prototypes can also be observed through the changes of cluster
properties for the hit clusters produced by the beam. Figure \ref{CS_vs_resistivity} shows the typical cluster size measured for the high-gain matrix, that shows the highest detection efficiency. For all resistivities, the dataset is dominated by clusters containing only one pixel. This is due to the large size of the pixel (\SI{250x50}{\um}) and the small depletion depth expected (\SI{<50}{\um}). At higher resistivity, the influence of a large depletion depth can be observed and two phenomena are competing. The charge deposited deeper in the bulk will drift for a longer time than the charge deposited close to the electrodes. Meanwhile, the increased electric field in the bulk increases the charge speed, making the charge drift time smaller. This second effect is enhanced in our prototype as the bias voltage is applied on an electrode surrounding the collection diode. As the depletion depth reaches values larger than the electrode to N-Well distance (\SI{10}{\um}), the behaviour of the diode diverges from that of a planar diode. It can be observed in figure \ref{8c} that the charge sharing producing larger clusters is maximal at an intermediate bias voltage of \SI{80}{\V}.

\begin{figure}[!htbp]
\center{\includegraphics[width=0.7\textwidth]{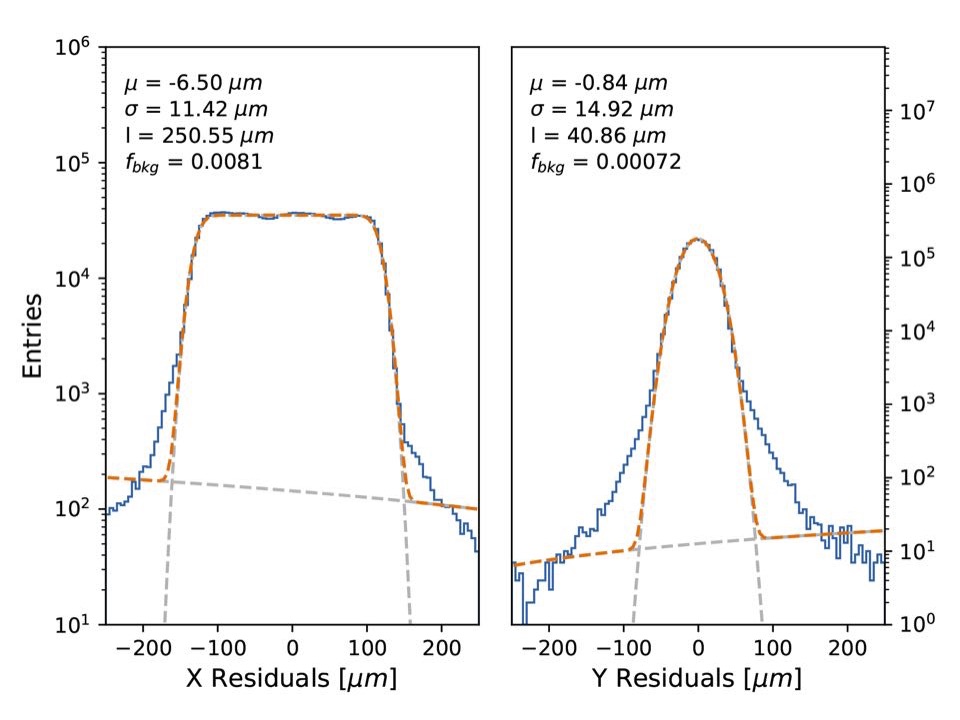}}
\vspace*{-0.4cm}\caption{Unbiased residual for a \SI{200}{\ohm\cm} sample, \SI{160}{\V}
  bias voltage, \SI{2000}{e} threshold, analog matrix 1 (blue). Fit of a gaussian convoluted with a box function (orange). $\mu$ is the mean of the distribution, $\sigma$ the width of the gaussian, $l$ the width of the box and $f_{bkg}$ the fraction of background events.}
  \label{residuals}
\end{figure}

As most clusters contain only one pixel, the spatial resolution of the
prototypes is mainly determined by the size of the pixels. Figure
\ref{residuals} shows a typical unbiased residual distribution for a
\SI{200}{\ohm\cm} sample operated with \SI{160}{\V} bias voltage and
\SI{2000}{e} threshold. Some tails can be observed on the residual
curves. These can be explained as clusters containing more than one pixel,
with the charge induced due to capacitive coupling between a pixel pad
and its neighbours, as previously observed \cite{ccpdv3}. This charge is
then due to the coupling method and not due to the charge sharing inside
the bulk of the sensor. The cluster position is reconstructed using the
time-over-threshold-weighted center of gravity. The cross-coupling
increases the measured time-over-threshold away from the true hit
position. Consequently, the reconstructed cluster position is also
calculated to be further away and the residuals are enlarged. As the particle beam illuminating the sensor is not uniform, the fraction of background events is dependent on the sensor, resulting in the tilt observed on the residual plots.  

\begin{figure}[!htbp]
\centering
  \subfigure[\SI{80}{\ohm\cm}]{\includegraphics[width=0.49\textwidth]{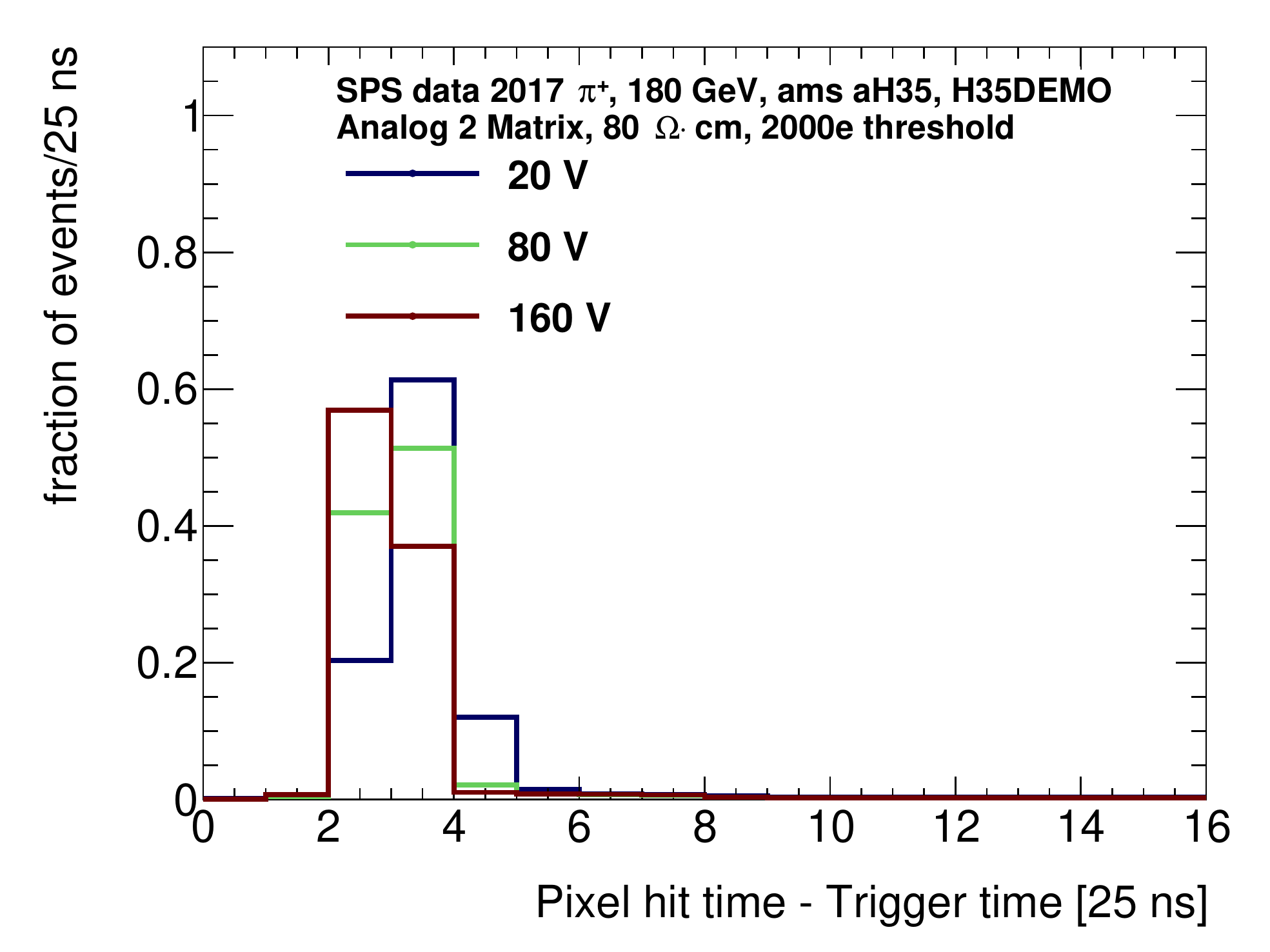}}
\hfill
  \subfigure[\SI{200}{\ohm\cm}]{\includegraphics[width=0.49\textwidth]{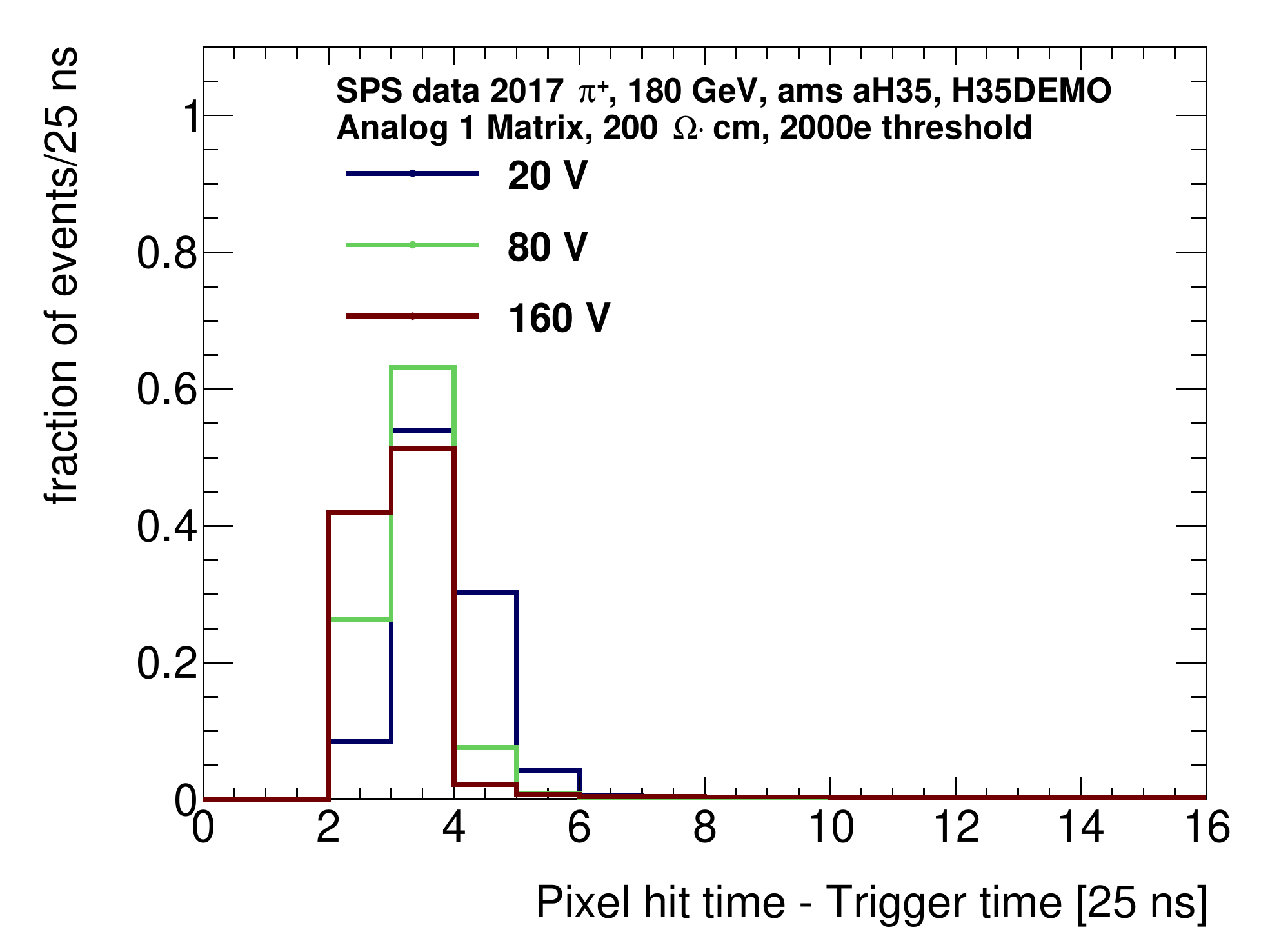}}\newline
\hfill     
  \subfigure[\SI{1000}{\ohm\cm}]{\includegraphics[width=0.49\textwidth]{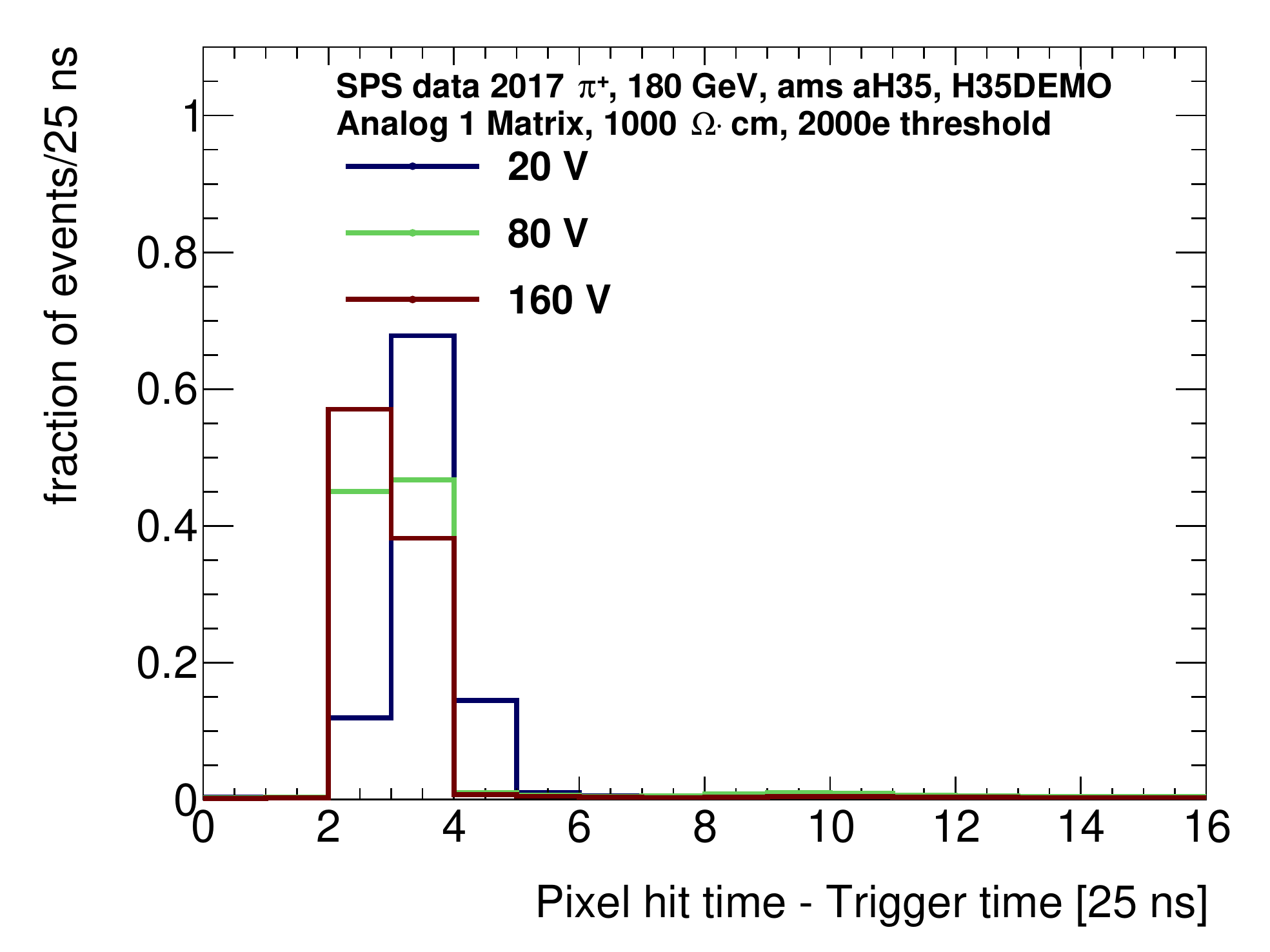}}
  \caption{Signal arrival time distributions in \SI{25}{\ns} bins versus
    bias voltage and different substrate resistivity for \SI{2000}{e}
    threshold.}
  \label{timing_vs_resistivity}
\end{figure}

The resistivity of the substrate should also have an effect on the rise time of the signal and the amount of charge generated. Figure \ref{timing_vs_resistivity} shows the dependence of the timing resolution of the H35DEMO for different bias voltage and resistivity for the second high-gain matrix. No significant variation of the distribution was observed in the other matrices. It was not possible to adjust the delay of the clock for the FE-I4 ASIC to optimise the binning of the timing distribution. No clear dependence of the timing resolution on resistivity, gain or presence of the Deep P-Well can be observed. However, in all cases, the timing distribution is constrained to less than \SI{50}{\ns} for bias voltages over \SI{80}{\V}. We can deduct from this that the timing resolution is dominated by the intrinsic jitter of the preamplifier and not by the sensor signal amplitude or rise time.

\subsection{Detection efficiency}

\begin{figure}[!htbp]
  \subfigure[Analog 1, submatrix 1]{\includegraphics[width=0.5\textwidth]{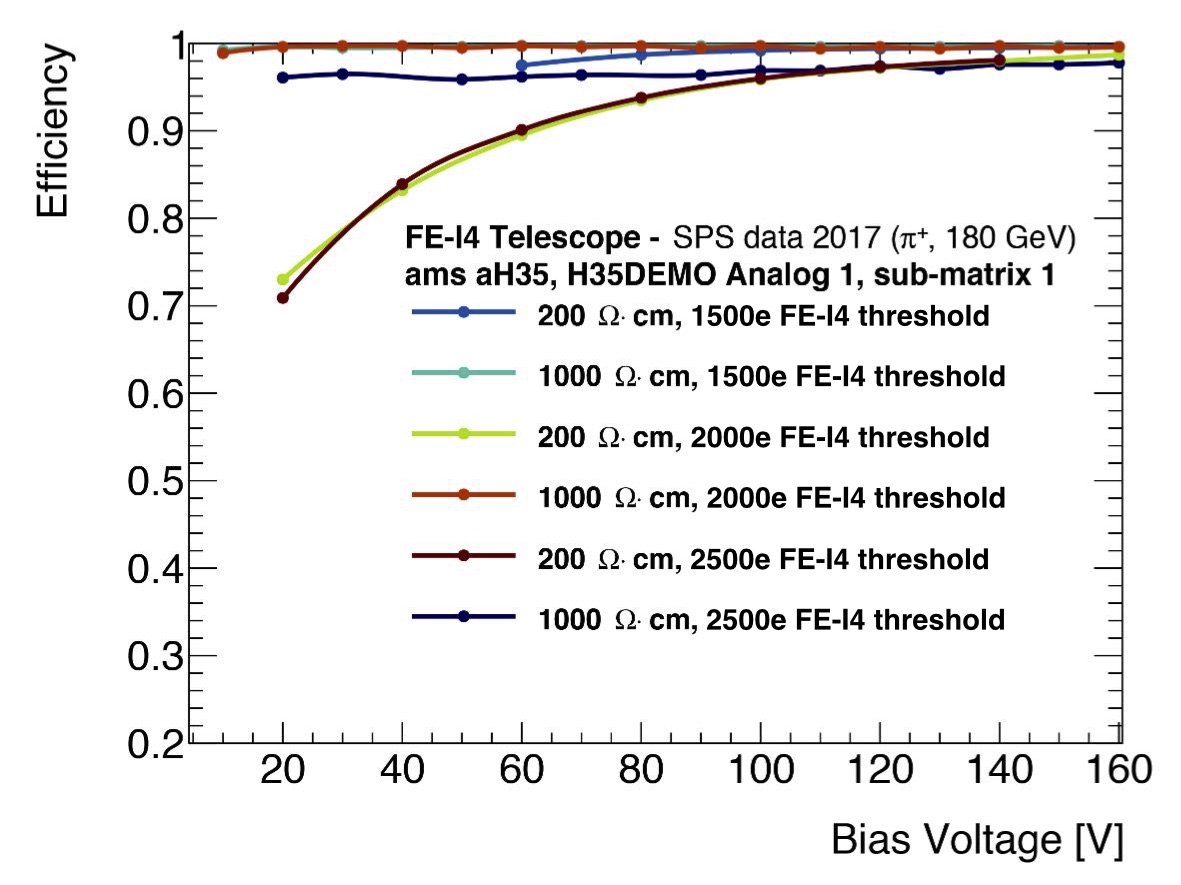}}
  \subfigure[Analog 2, submatrix 1]{\includegraphics[width=0.5\textwidth]{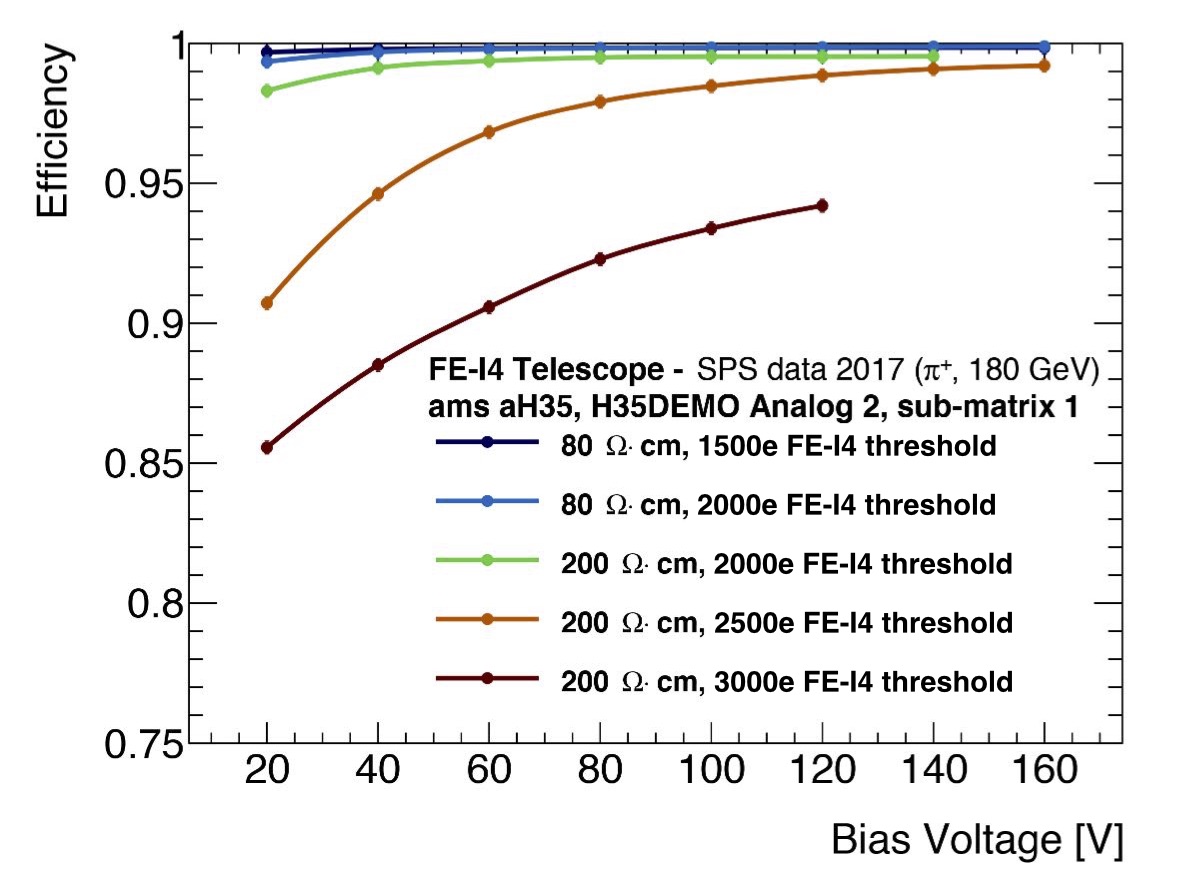}}      
  \subfigure[Analog 1, submatrix 2]{\includegraphics[width=0.5\textwidth]{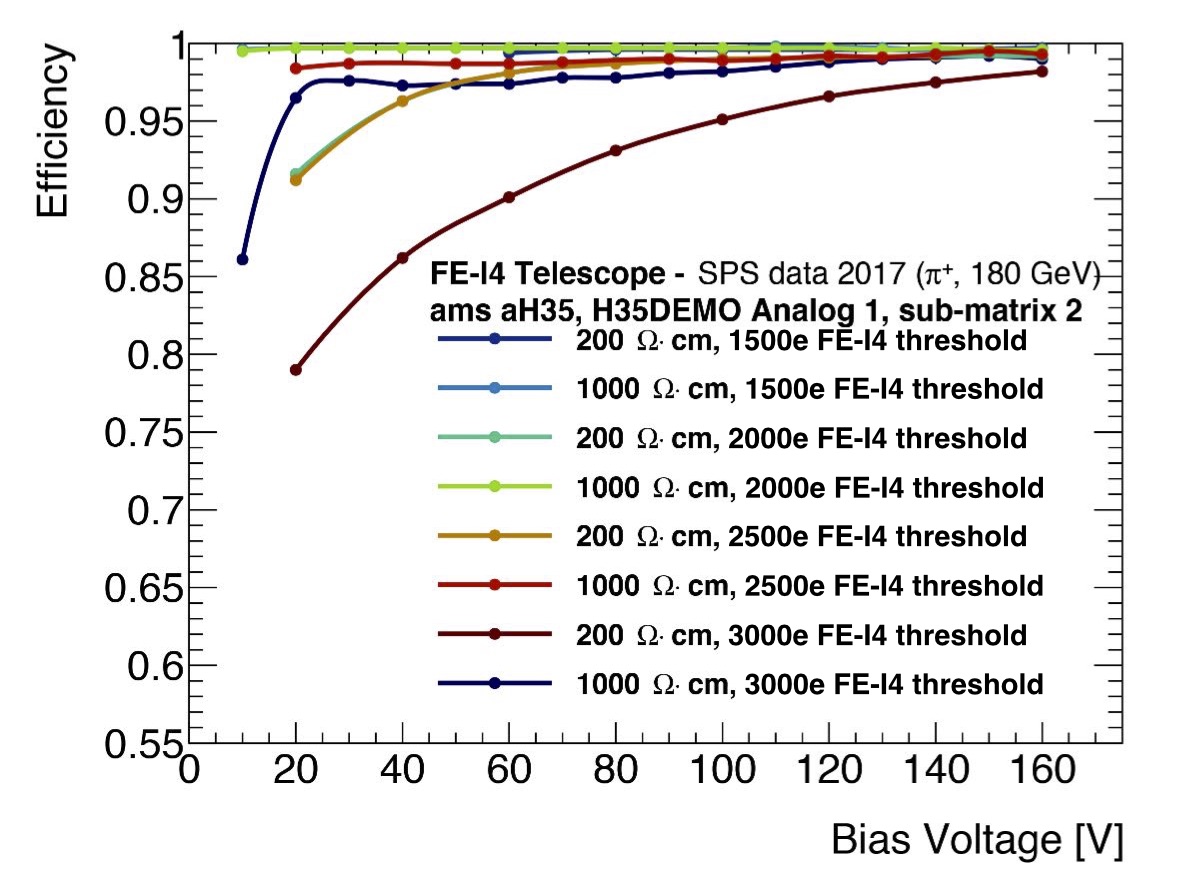}}
   \subfigure[Analog 2, submatrix 2]{\includegraphics[width=0.5\textwidth]{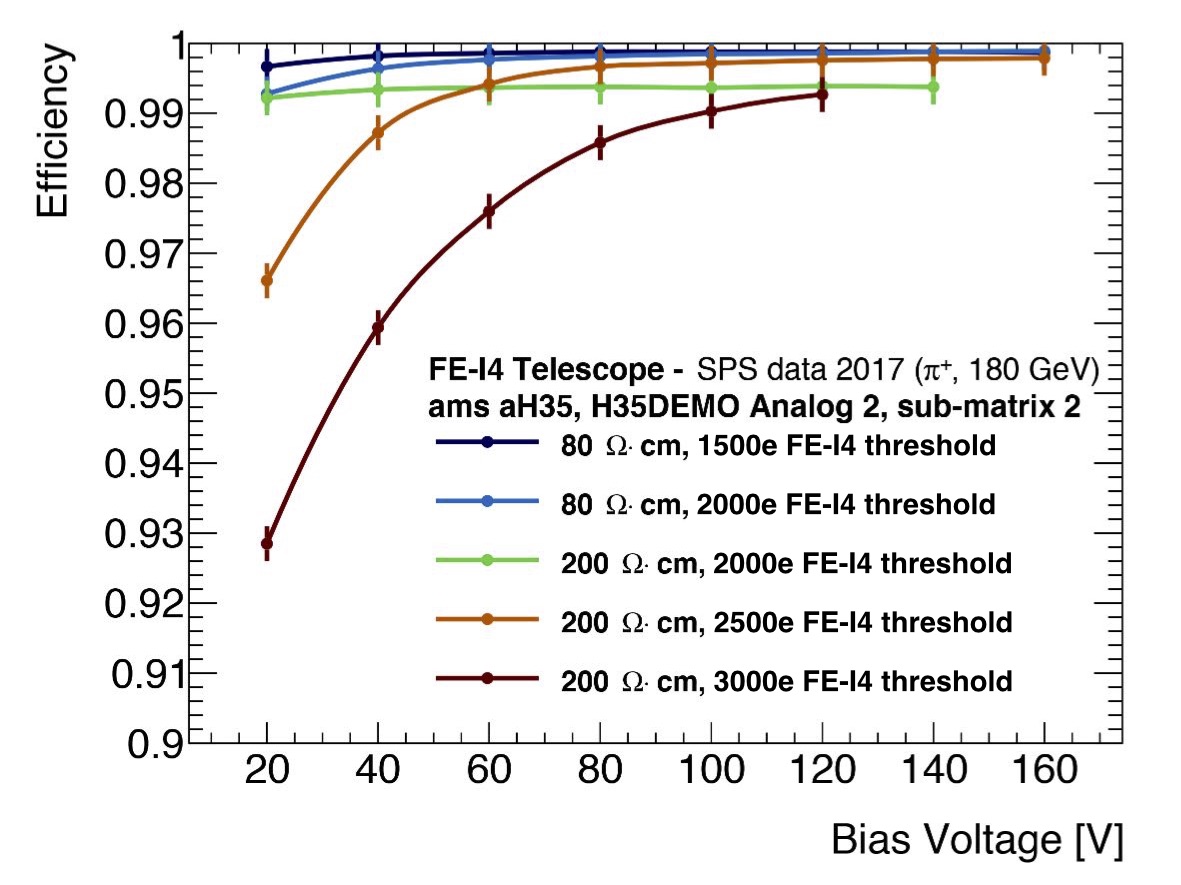}}  
   \subfigure[Analog 1, submatrix 3]{\includegraphics[width=0.5\textwidth]{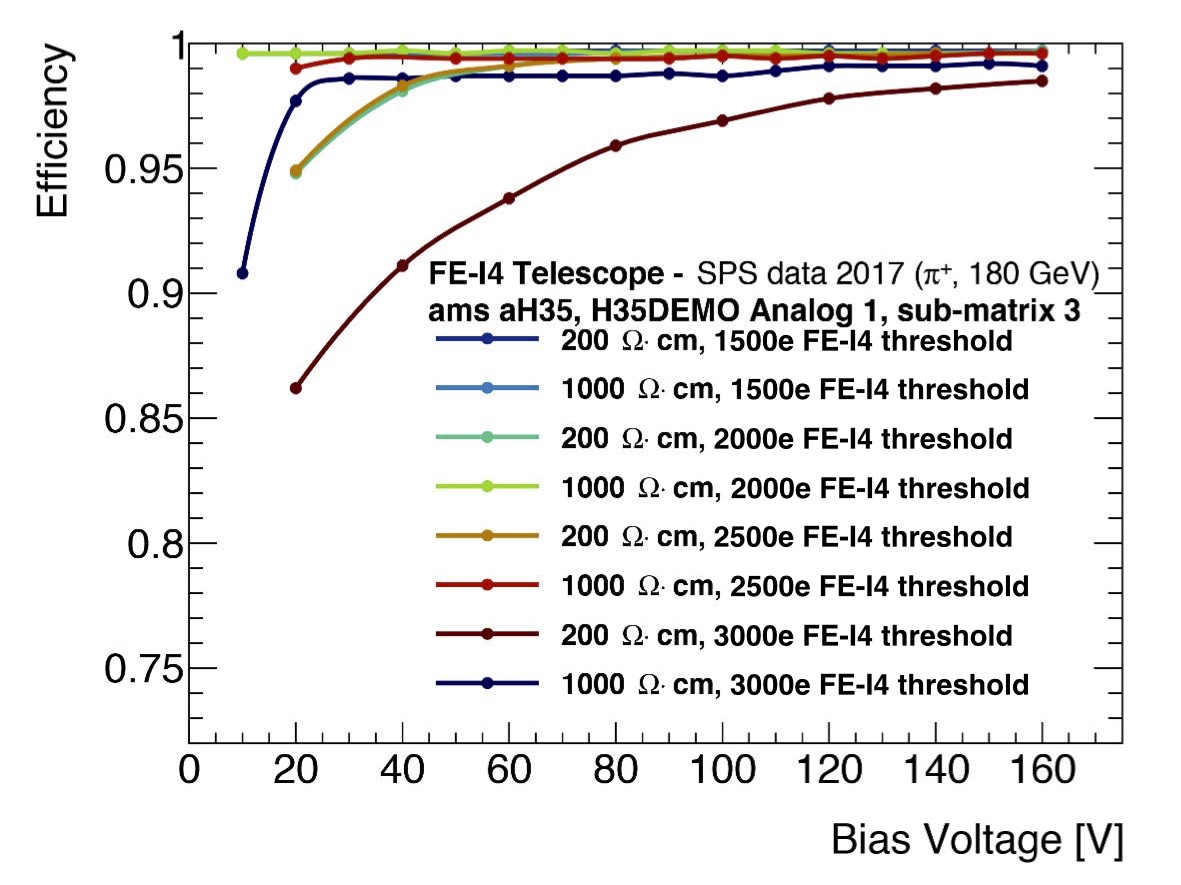}}  
  \subfigure[Analog 2, submatrix 3]{\includegraphics[width=0.5\textwidth]{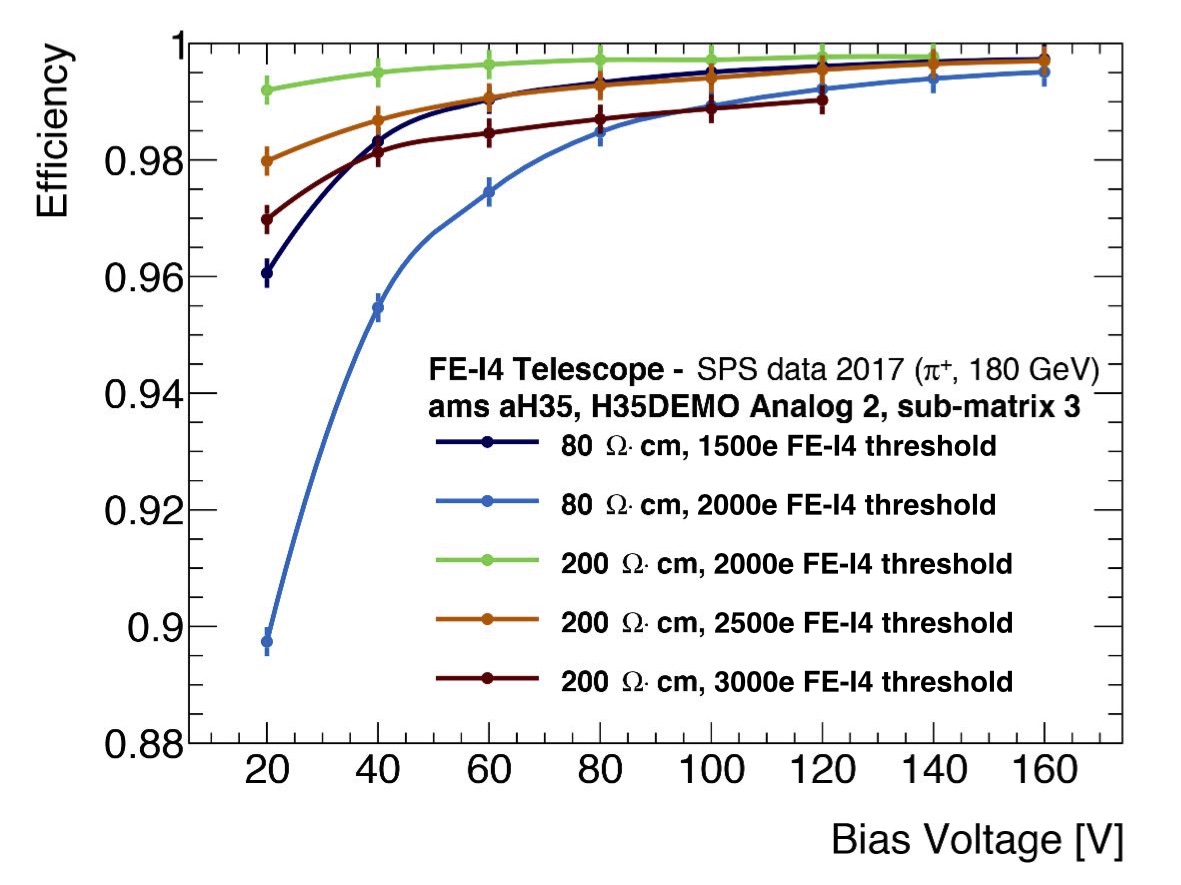}}   
  \caption{Global efficiency of \SIlist{80;200;1000}{\ohm\cm} resistivity substrate samples for the different sub-matrices as a function of bias voltage for different substrate resistivities and FE-I4 detection threshold.}
  \label{Eff_vs_resistivity}
\end{figure}

The particle detection efficiency is a key parameter to determine the
usability of high-resistivity substrate CMOS sensors. This parameter is
influenced by the signal strength and the gain of the
preamplifier. Figure \ref{Eff_vs_resistivity} shows the threshold scan
performed for the individual matrices for different thresholds settings of the
FE-I4. A clear dependence of the efficiency on the substrate resistivity
for comparable thresholds can be observed. The higher the resistivity,
the smaller the bias voltage needed to obtain a good detection
efficiency, in agreement with our expectation. Excellent detection
efficiency superior to \SI{99}{\%} can be achieved for all resistivities
studied with a threshold of \SI{2000}{e}, corresponding to approximately
\SI{1500}{e} signal in the H35DEMO sensor, assuming
\SI{3.5}{\femto\farad} coupling capacitance and a gain of
\SI{100}{\mV\per{1500e}}.

\begin{figure}[!htbp]
\centering
  \subfigure[\SI{80}{\ohm\cm}, \SI{160}{\V}, \SI{2000}{e} ]{\includegraphics[width=0.49\textwidth]{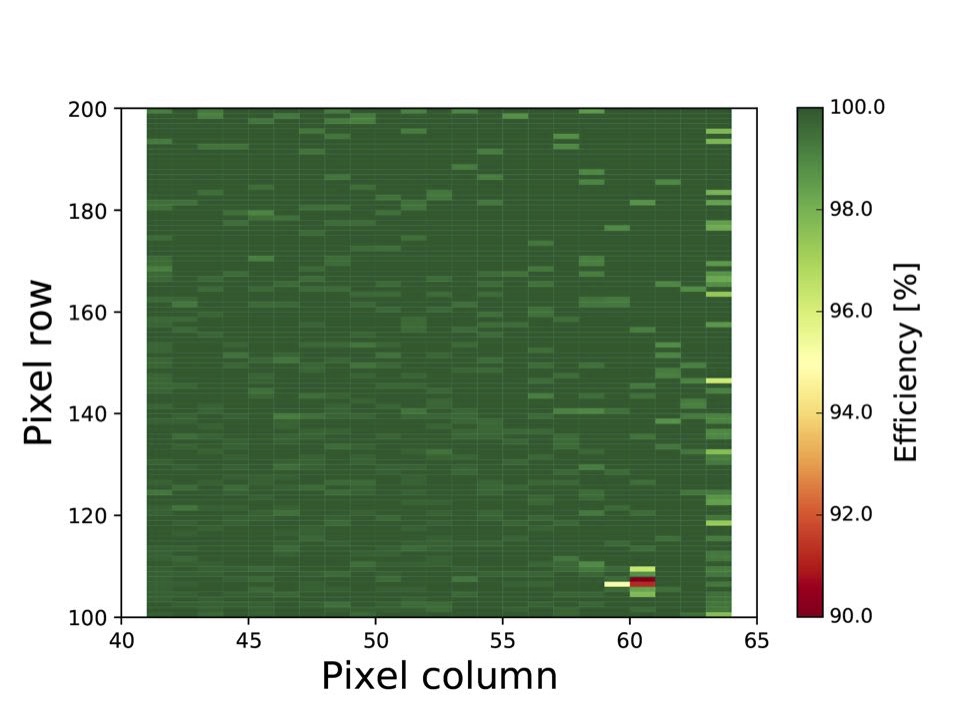}}
\hfill
  \subfigure[\SI{200}{\ohm\cm}, \SI{160}{\V}, \SI{2250}{e}]{\includegraphics[width=0.49\textwidth]{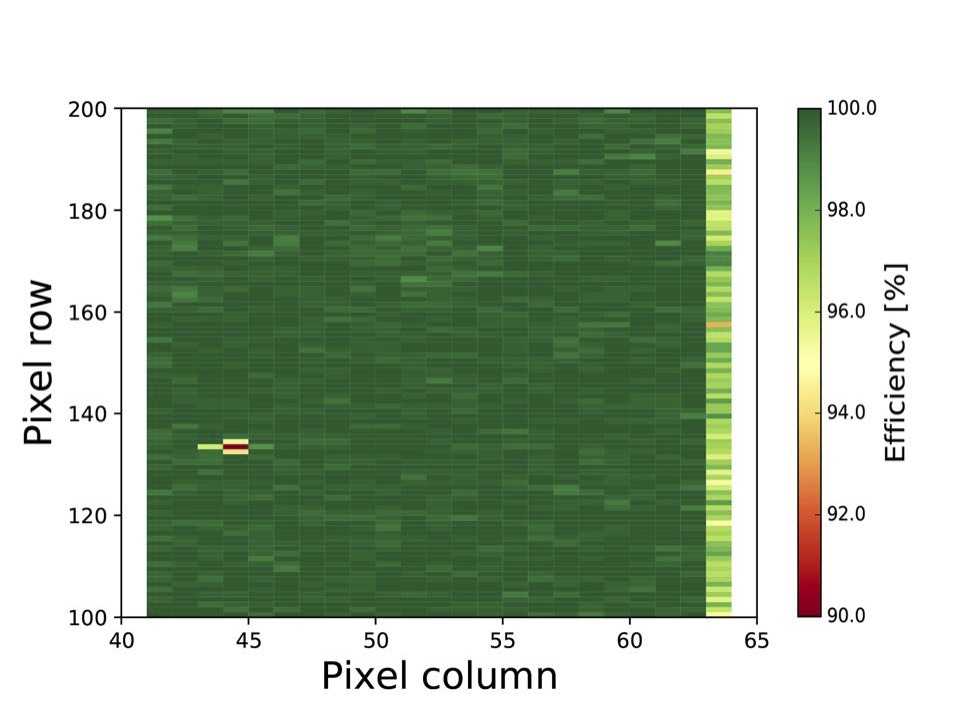}}\newline
\hfill     
  \subfigure[\SI{1000}{\ohm\cm}, \SI{160}{\V}, \SI{2000}{e}]{\includegraphics[width=0.49\textwidth]{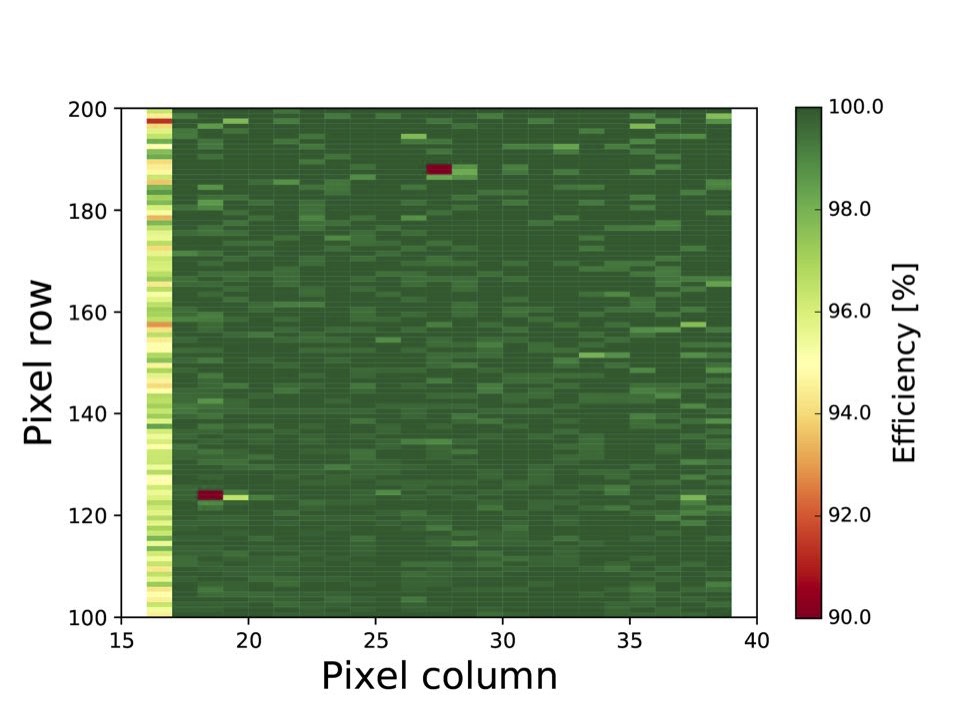}}
  \caption{Efficiency for the analog high-gain sub-matrices, covering rows 100 to 200 of the FE-I4 and spanning over 24 columns, at \SI{160}{\V} and different substrate resistivity for \SI{2000}{e} threshold. The left column for analog matrix 2 and right column for analog matrix 1 show lower efficiency due to the proximity with the pixels of the monolithic NMOS and CMOS matrices, which can compete for the charge signal.}
  \label{GlobalEfficiency_vs_resistivity}
\end{figure}

Figure \ref{GlobalEfficiency_vs_resistivity} shows the efficiency for
each pixel of the second high-gain matrix for each of the three
resistivities, taken at \SI{2000}{e} FE-I4 threshold and \SI{160}{\V} bias
voltage. These results show, after careful tuning of the glueing method, a good detection uniformity over the matrix has been achieved. This demonstrates a good uniformity of the pixel's electrical behaviour within the columns and good uniformity of the glue interface and preamplifier properties.

\begin{figure}[!htbp]
\center{
  \subfigure[\SI{80}{\ohm\cm}, \SI{20}{\V}, \SI{2000}{e} threshold]{\includegraphics[width=0.45\textwidth]{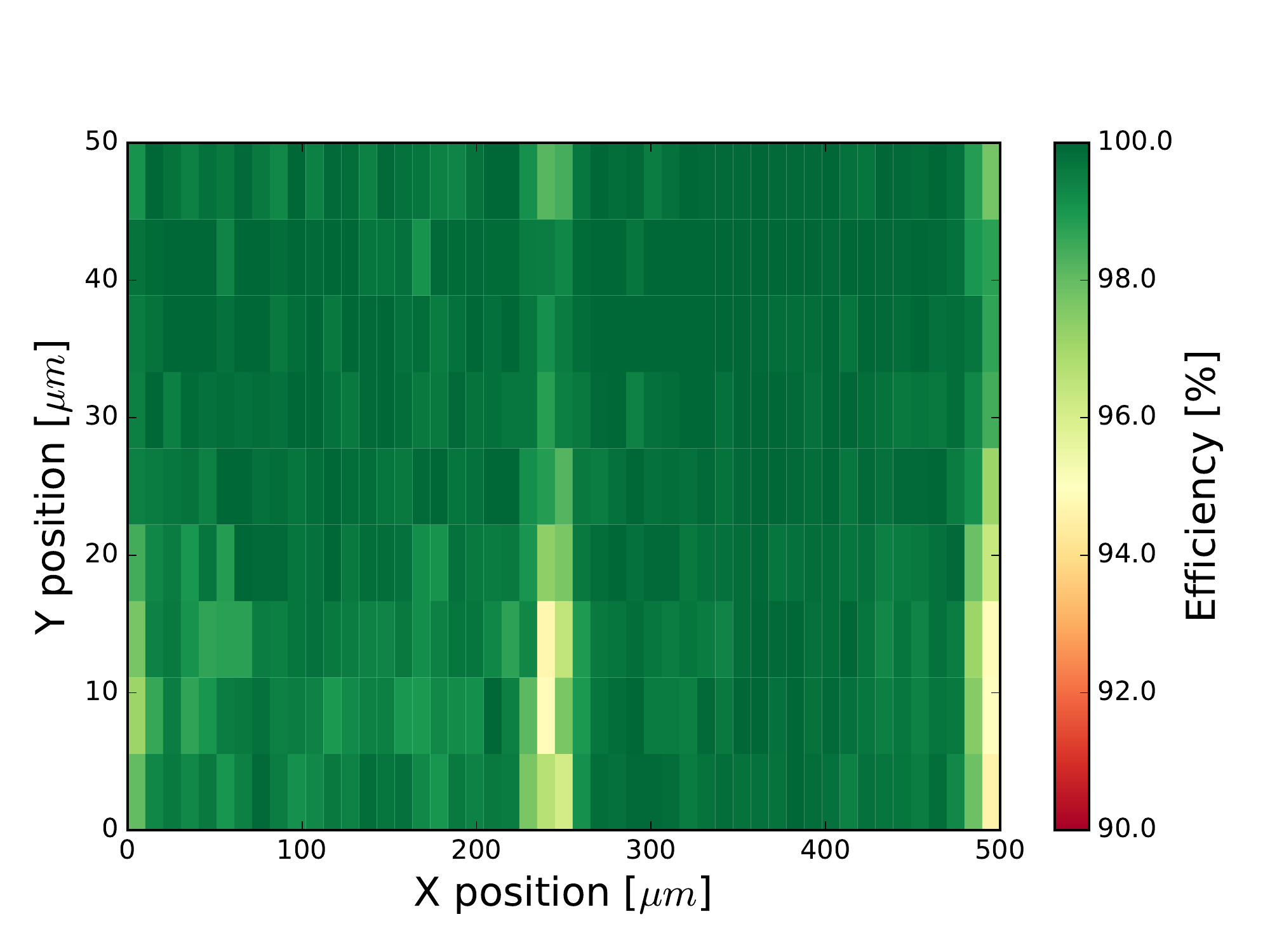}}\label{lowinpixeff}
  \subfigure[\SI{80}{\ohm\cm}, \SI{160}{\V}, \SI{2000}{e} threshold]{\includegraphics[width=0.45\textwidth]{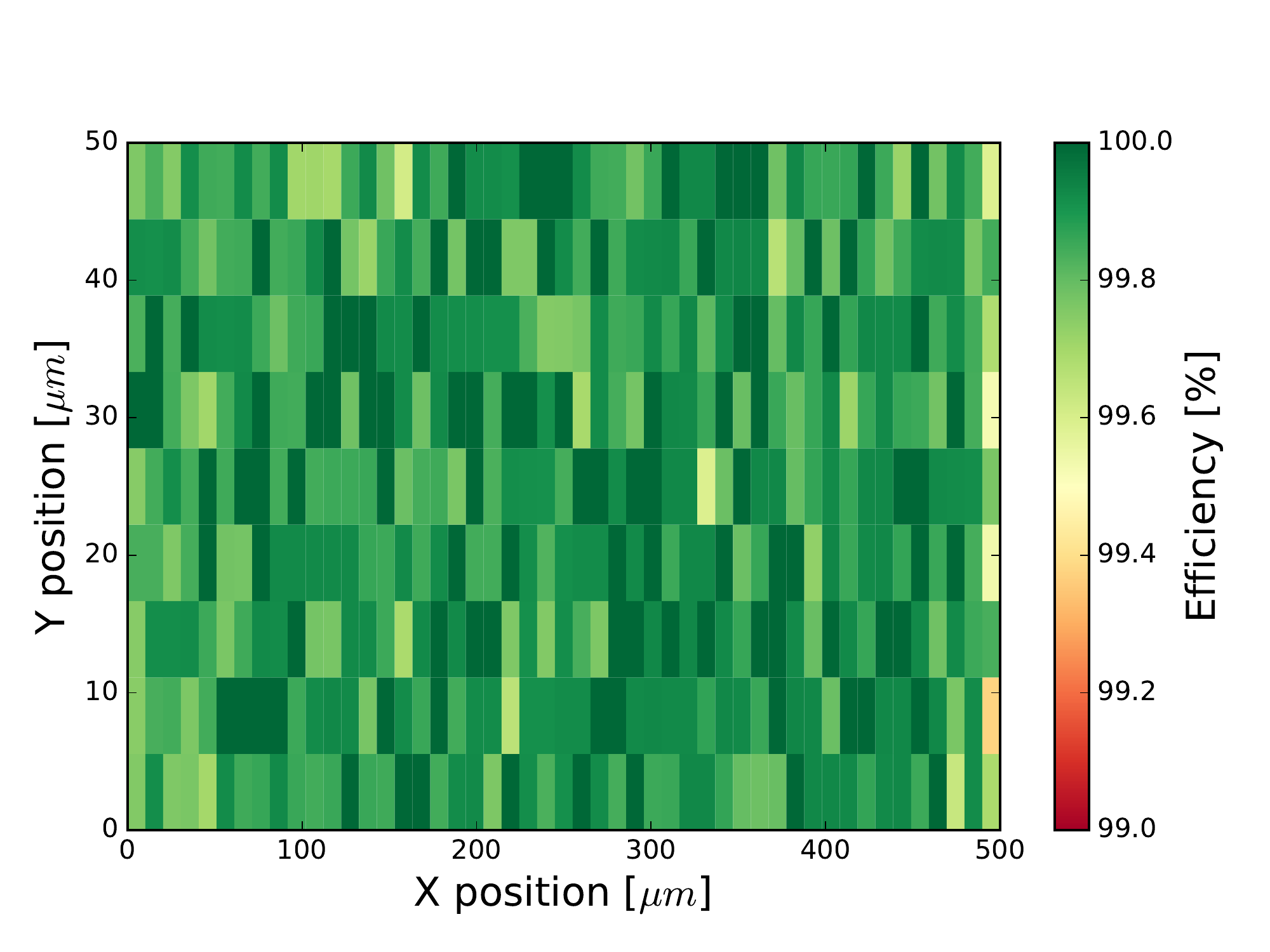}}
  \subfigure[\SI{200}{\ohm\cm}, \SI{20}{\V}, \SI{2000}{e} threshold]{\includegraphics[width=0.45\textwidth]{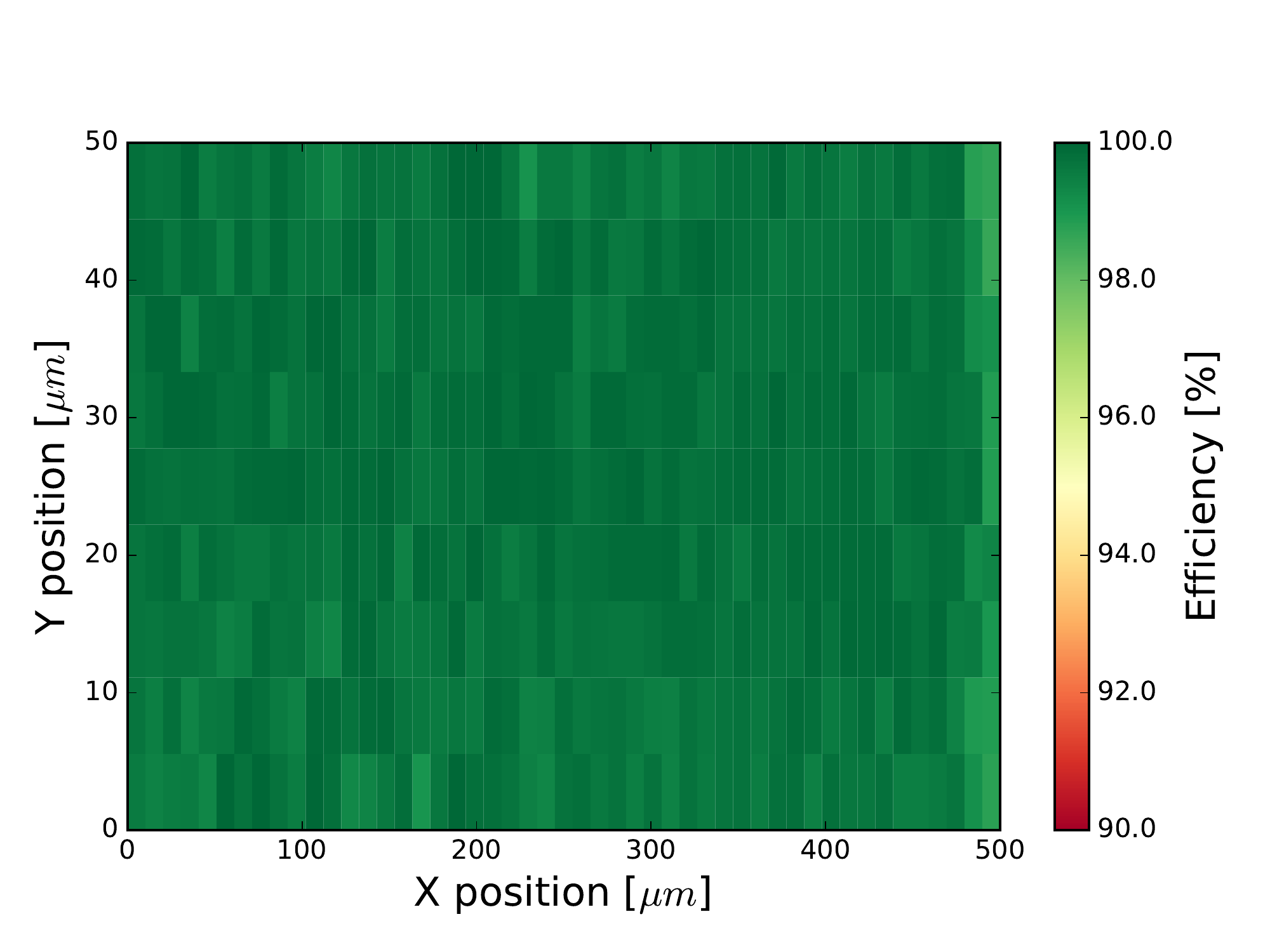}}
   \subfigure[\SI{200}{\ohm\cm}, \SI{160}{\V}, \SI{2250}{e} threshold]{\includegraphics[width=0.45\textwidth]{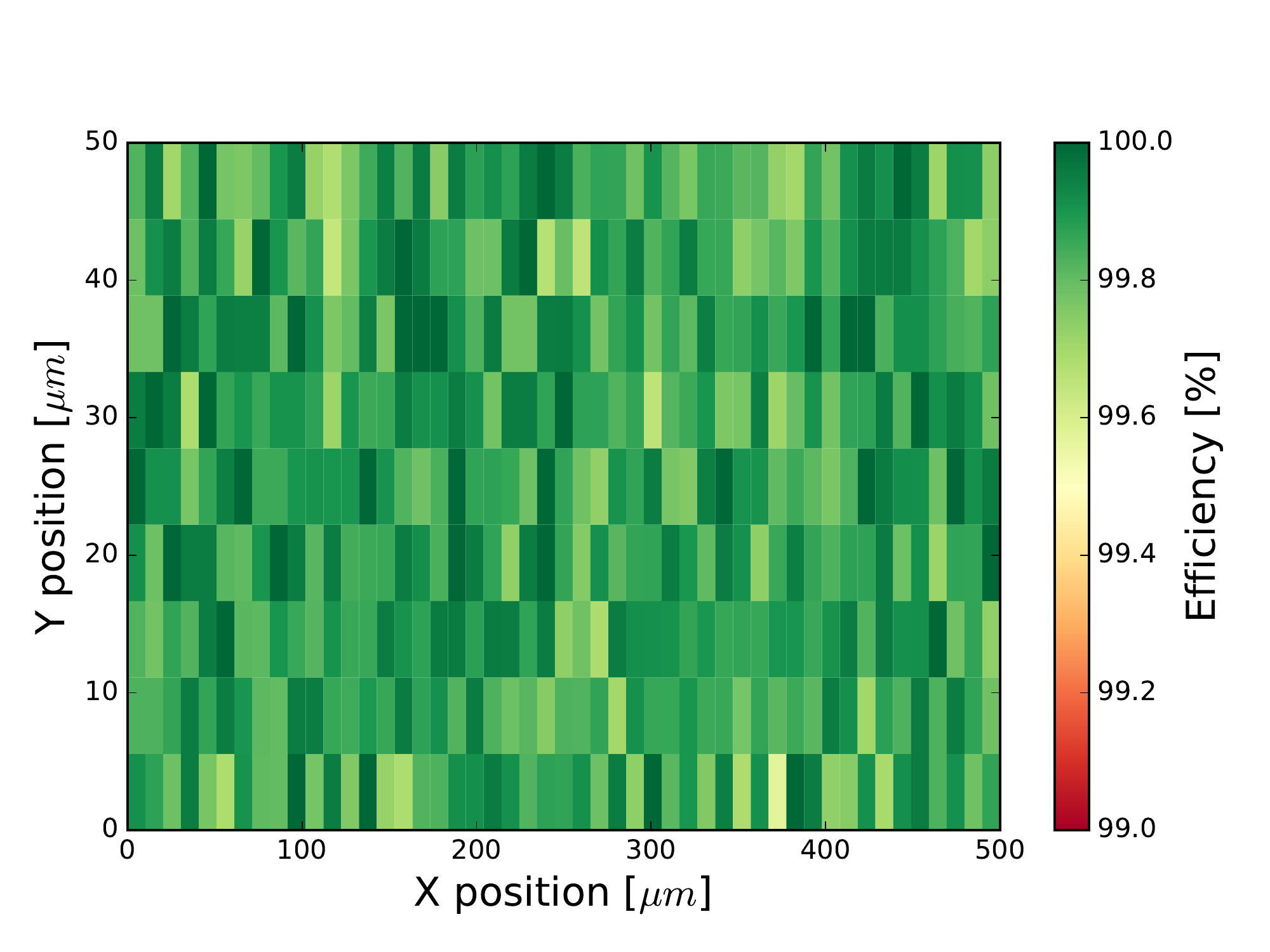}}
   \subfigure[\SI{1000}{\ohm\cm}, \SI{10}{\V}, \SI{2000}{e} threshold]{\includegraphics[width=0.45\textwidth]{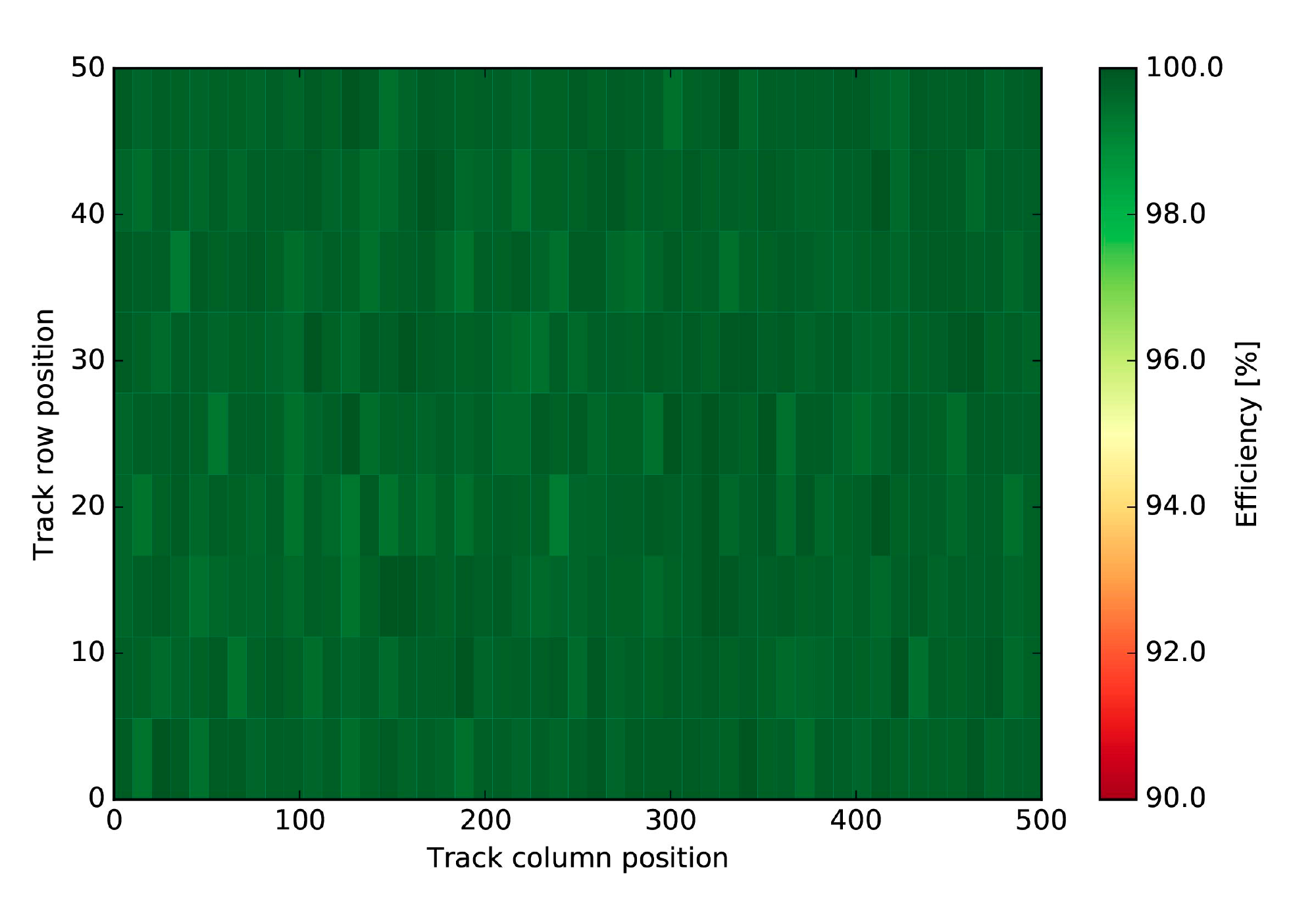}}
  \subfigure[\SI{1000}{\ohm\cm}, \SI{160}{\V}, \SI{2000}{e} threshold]{\includegraphics[width=0.45\textwidth]{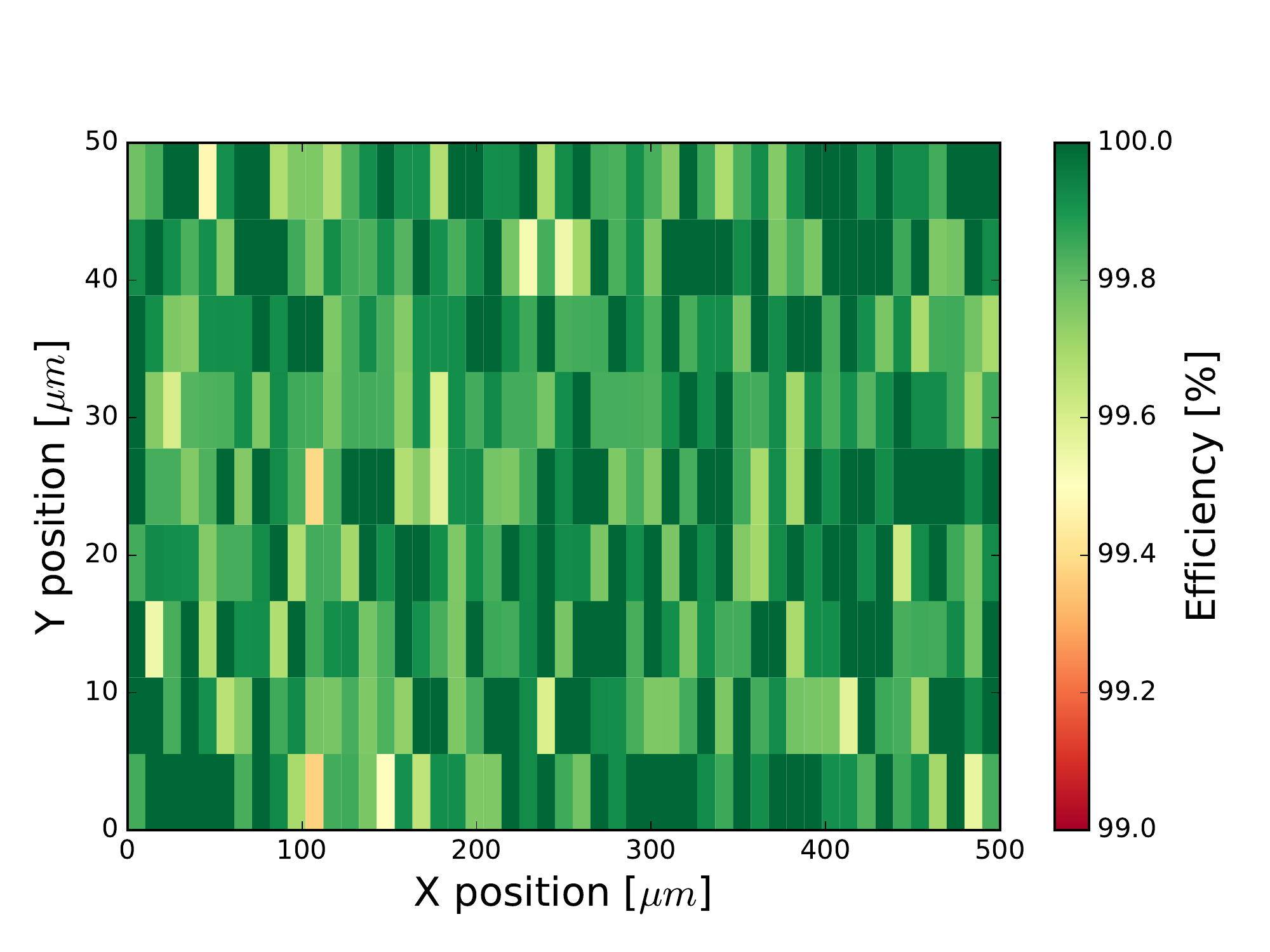}}
  \caption{In-pixel efficiency of \SIlist{80;200;1000}{\ohm\cm} resistivity substrate samples for sub-matrix 2 (Analog 2 for 80 and \SI{200}{\ohm\cm}, Analog 1 for \SI{1000}{\ohm\cm}). Pixels are arranged in a symmetric double column pattern. Note the difference in the color scale between data at \SI{160}{\V} and \SI{20}{\V} and the different threshold for d). }
  \label{Effinpix_vs_resistivity}
  }
\end{figure}

Figure \ref{Effinpix_vs_resistivity} shows the in-pixel efficiency measured for the high-gain matrices for different resistivities and HV bias. Figure \ref{Effinpix_vs_resistivity} a), c) and e) shows the efficiency at low bias voltage, where intra-pixel regions with lower efficiency are visible, more evident for the lower resistivity substrate. TCT measurements \citep{tct} has confirmed that lower depletion volumes are achieved with lower resistivity substrates, when compared with an higher resistivity substrate at the same HV bias. This effect can be observed as the less efficient region between pixels as shown on Figure \ref{lowinpixeff}. All samples could be operated with high efficiency when sufficient bias was applied, as shown on figure \ref{Effinpix_vs_resistivity} b), d) and f).

In light of the results obtained during this test beam campaign, the following observations can be made  : 
 \begin{itemize}
   \itemsep0em 
   \item The use of ELT in the feedback circuitry does not affect significantly the performance of the amplifier, as shown on figure \ref{ana1}.
  \item The extra deep P-Well (DPTUB) should not be used, placed under the P-Well providing the contact to the substrate, in order to reduce the input capacitance of the preamplifier. Figure \ref{submatrix_eff} shows that a higher efficiency is achieved without the DPTUB.
  \item High-Gain is required to achieve good detection efficiency over a large range of bias in these conditions, as results from the second analog matrix has shown.
  \item Time resolution of the sensor is limited by the amplifier power consumption but is not affected significantly by the gain or feedback transistor used. The H35 technology uses 3.3V power supplies for the front-end and the current distributed to the pixels was limited to allow for efficient cooling during operation. A higher power consumption or a transition to a 1.8V technology would yield to better timing, as show in our previous results with the IBM h18 technology \cite{CCPDv4-paper,CCPDV4irrad}.
  \item A threshold of \SI{1500}{e} or less must be achieved to ensure good detection efficiency superior to \SI{99}{\%} for all resistivities when the discriminator is to be implemented in the pixel, for monolithic integration (\SI{2000}{e} equivalent in FE-I4).
\end{itemize}

\section{Conclusion}

Extensive test beam measurements of capacitively coupled pixel detectors designed in ams aH35 HV-CMOS technology were performed to evaluate the effects of high-resistivity substrates on the properties of the detector and evaluate the feasibility of building large area CMOS sensors. A method for uniform and reproducible glueing of the H35DEMO prototypes to the FE-I4 ASIC was developed and successfully used to produce a series of prototypes in three resistivities that were measured in beam tests. The results of this investigation shows that detection efficiencies larger than 99\% can be achieved for all prototypes. A good uniformity in the coupling and on detection efficiency over a large area was measured. The advantages of using higher resistivity are illustrated by the measurement of the detection efficiency as a function of bias voltage and FE-I4 threshold. Higher resistivity results in larger signals and better efficiency at lower bias voltage. The time resolution of the different prototypes were evaluated and the results show little dependence on the substrate. This indicates that the timing resolution is determined by the preamplifier itself, with little influence from the signal strength and varying rise time. %The next steps of this study will be to study the same tracking properties for irradiated sensors where phenomena such as acceptor removal \cite{acceptorremoval} will increase the depletion in the $80$ and $200~\Omega\cdot cm$ samples while decreasing the depletion in the $1000~\Omega\cdot cm$.

\acknowledgments
The authors gratefully acknowledge the support by the
CERN PS and SPS instrumentation team and Fermilab Test Beam Facilities.
The research presented in this paper was supported by the SNSF grants
20FL20\_173601,200021\_169015 and 200020\_169000 and by the H2020 project AIDA-2020, GA no. 654168.

\end{document}